# Defect Engineering of Ultrathin Gallium Nitride *via* Electric Fields for Advanced Electronic, Magnetic, and Gas Sensing Applications

Yujia Tian[a, b], Devesh R. Kripalani[a], Ming Xue[b], and Kun Zhou[a, c*]

[a]School of Mechanical and Aerospace Engineering, Nanyang Technological University, 50 Nanyang Avenue, Singapore 639798, Singapore

[b]Infineon Technologies Asia Pacific Pte. Ltd., 168 Kallang Way, Singapore 349253, Singapore

[c]CSIJRI-Qingyang Innovation Joint Lab, China-Singapore International Joint Research Institute, Guangzhou 510700, China

[*]Email: kzhou@ntu.edu.sg

## Abstract

Scaling wide-band-gap semiconductors to the ultrathin limit offers a transformative pathway for power electronics, with gallium nitride (GaN) representing a cornerstone material in this class. However, the operational resilience and functional tunability of its two-dimensional form (g-GaN) remain underexplored. This work shifts the focus from idealized systems to the complex materials behavior under realistic conditions, investigating how the synergistic effects of point vacancy defects, strain, and external electric fields govern its electronic, magnetic, and sensing landscapes. We demonstrate that these factors are not merely perturbations but are fundamental to modulating the material response. Our first-principles calculations suggest g-GaN maintains electronic stability under intense electric fields; notably, gallium vacancies are predicted to further extend the theoretical stability limit. While in-plane tension preserves the band gap evolution under an electric field, in-plane compression facilitates low-field metallization. Using nitrogen monoxide (NO) adsorption as a prototype, we find that the interaction is defect-modulated and potentially tunable by electric fields. Analysis of adsorption energetics and diffusion barriers suggests the gallium vacancy may act as a thermodynamic trap for NO. Targeted hybrid-functional (HSE06) validation confirms the reliability of observed adsorption trends and theoretical metallization thresholds, while revealing that precise electronic-exchange treatment is critical for capturing the magnetic ground state of nitrogen vacancies. By systematically examining the geometry, energetics, band structure, density of states, magnetic response, and charge transfer, this study clarifies the interplay between defects and external electric fields, providing insights into theoretical upper bounds for property tuning and semiconductor device engineering.

## Introduction

Gallium nitride (GaN) is a pivotal wide-band-gap semiconductor that revolutionized optoelectronics through its use in blue light–emitting diodes, a breakthrough recognized by 2014 Nobel Prize in Physics for enabling energy-efficient lighting and displays.[1] Beyond the foundational role, the exceptional electronic and thermal properties of GaN drive innovations in high-power transistors,[2-4] high-frequency wireless communication,[5] renewable energy systems,[6, 7] and water splitting.[8, 9] As device miniaturization demands nanoscale solutions, two-dimensional (2D) GaN with atomic-scale thickness emerges as a transformative frontier. It exhibits enhanced electron mobility,[10] low thermal conductivity for thermoelectrics,[11] and versatile capabilities of heterogeneous integration,[12] while its expansive surface area unlocks potential in gas sensing.[13-21] These attributes position ultrathin GaN as a critical material for next-generation nanoelectronics and energy technologies, urging deeper exploration of its physics.

While bulk GaN adopts a nonlayered wurtzite structure,[22] its transformation into 2D GaN destroys the Ga–N bond along the *c* axis and $sp^3$ orbital hybridization in the original $GaN_4$ tetrahedron, replacing the buckled configuration with a planar, graphene-like phase (g-GaN) of $D_{3h}$ symmetry.[23] The structural shift reconfigures the bonding and electronic behavior, endowing g-GaN with properties distinct from its bulk counterpart. This honeycomb phase was first proven to be stable by density functional theory (DFT) calculations in 2009.[24] Thereafter, more experimental studies followed and achieved efficient fabrication of 2D GaN in various forms,[10, 25-27] including a free-standing sheet.[28] Yet, despite the demonstrated stability of this atomically thin phase, the fundamental mechanisms governing its property modulation remain underexplored, especially under the synergy between real-world factors. They are essential to unlocking its full potential, where its ultrathin geometry and tunable properties could redefine device performance limits.

Being a wide-band-gap semiconductor, GaN is regarded as a robust material for high-power applications, which require strong electric fields. Electric fields may also be used to regulate the spin-related electronic structures of 2D materials for spintronic and multistate information storage.[29, 30] The rapid nature of this property modulation can induce high-frequency dynamic switching in 2D materials, which can revolutionize their responsiveness. However, the electric field–dependent ferromagnetic behavior of atomically thin materials may be difficult to realize because of their intrinsic inversion symmetry. It causes a symmetric spatial distribution of charge density that makes the material unresponsive to spontaneous polarization from external electric fields. Defects may be deliberately introduced in 2D materials to overcome such constraints posed by symmetry and tune electronic properties.[31-37] They may also form easily and unintentionally during material fabrication. Particularly, point defects, as dictated by the second law of thermodynamics, are inherent to all materials and are especially noticeable in chemically grown 2D materials because of imperfections in the growth process. While previous work demonstrated the potential of defect engineering in g-GaN,[38-40] for the promising material to be fully exploited for electronic and memory storage applications, especially those requiring high voltage, the concurrent impact of electric fields cannot be ignored.

Herein, we systematically unravel the promising potential of g-GaN in its defect- and electric field–tunable behavior through a hierarchical first-principles exploration. It begins with studying the properties of pristine g-GaN for setting the stage for defect engineering. Gallium ($V_{Ga}$) and nitrogen ($V_N$) vacancies are next shown to reconfigure the electronic and magnetic behavior with atomic precision. The concurrent effect of strain is also considered to reveal their easy introduction. Using nitrogen monoxide (NO) as a model adsorbate, we demonstrate how vacancies may act as atomic-scale traps for detection and removal of gases, while the same gas can also be exploited for defect detection in the material. Finally, we unveil the synergy between vertical electric fields, strain, defects, and gas interaction, demonstrating how field-driven structural and electronic reconfigurations can enable on-demand transitions between semiconducting/metallic states. From fundamentals of the pristine material to defect-enabled innovation and adaptive field control, this cascading narrative not only deciphers the atomic-scale mechanisms governing the versatility of g-GaN but also charts a roadmap for defect-engineered 2D materials in next-generation nanoelectronics and environmental sensors.

## Results/Discussion

### *Comparison between pristine and defective g-GaN*

The pristine unit cell structure of g-GaN is optimized first *via* DFT calculations (Figure S1a). It has a lattice constant of 3.257 Å and bond length of 1.880 Å, consistent with literature values.[18, 39, 41-45] Its planar geometry is maintained by π bonds formed between the $p_z$ orbitals of gallium and nitrogen atoms.[46] Based on the high-symmetry path shown in Figure S1b, the band structure is computed and presented together with the density of states (DOS) in Figure S1c. Pristine g-GaN is a nonmagnetic semiconductor, and its indirect band gap between the K and Γ sites is calculated to be 1.94 eV, in good agreement with values reported in previous work.[44, 45] The valence band edge is mainly contributed by the p orbitals of nitrogen atoms, consistent with literature results.[47]

Crystalline defects are ubiquitous in 2D materials and can take various forms, such as grain boundaries,[48] dislocations,[49] antisite defects,[50] and vacancies.[51] Monovacancies often exhibit the lowest formation energy among atomic vacancies, as demonstrated in 2D molybdenum disulfide ($MoS_2$),[52] which highlights their easy formation and frequent occurrence. Advanced techniques such as ion or electron irradiation,[53] laser irradiation,[54] chemical etching,[55] and chemical vapor deposition (CVD) defect engineering [56] can provide precise control over the type, concentration, and spatial distribution of atomic vacancies, making them an attractive avenue for property tuning. For example, electron irradiation has been employed to precisely generate atomic defects in GaN.[57] By adjusting the acceleration voltage of the scanning electron microscope producing the electron beam, the defect concentration could also be tuned.[58]

To systematically explore the effects of vacancy defects in g-GaN, a 4 × 4 supercell is constructed for investigation. The same supercell is also used as the basis structure to study gas adsorption. With one gallium or nitrogen atom removed from the supercell to form either $V_{Ga}$ or $V_N$, respectively, the defect concentration is 6.25%. This percentage corresponds to a surface defect density of $6.8 \times 10^{13}$ cm$^{-2}$, which has the same order of magnitude as experimental values for films synthesized by CVD.[59-61] To decouple the fundamental magnetic and adsorption behavior from extrinsic charge correction effects, this work specifically examines the neutral case of the system. The geometry of the optimized structures with $V_{Ga}$ and $V_N$ defects is presented in Figure 1a and b, respectively. The $V_{Ga}$ defect causes the nitrogen atoms nearest to the vacancy to shift toward stable sites located slightly above and below the midplane of the monolayer, breaking the symmetry of the original configuration. In contrast, the structure with the $V_N$ defect remains planar. This difference indicates that only $V_{Ga}$ but not $V_N$ significantly alters the $\pi$ bonds formed by $p_z$ orbitals that maintain the planar geometry of g-GaN. The interatomic distance $d$ between gallium atoms closest to $V_N$ is 3.110 Å. It reduces by 4.5% from the pristine value, which could be attributed to the electrostatic attraction between the gallium atoms after removal of the nitrogen atom.

The formation energy $E_{form}$ of the $V_{Ga}$ and $V_N$ defects is 7.46 and 2.72 eV, respectively, consistent with literature values.[38-40] The significantly lower energy for $V_N$ indicates a higher likelihood of formation, potentially facilitated by the lighter atomic mass of nitrogen atoms. Considering that experimental 2D GaN sheets are prone to localized strain during exfoliation,[28] we examined the effect of lattice deformation and found that both compressive and tensile strains lower the formation barriers for these defects (by up to 10% at 2% biaxial strain, Figure S2), suggesting they may be even more prevalent in experimental samples. Healing of vacancy defects may be possible but only under conditions like high-temperature annealing,[62] which suggests the stability of the defective structures at room temperature.

The $V_{Ga}$ defect induces a striking reconfiguration to the band structure with spin-polarized bands near the Fermi level (Figure 1c). The vacancy introduces three unoccupied spin-down defect states within the band gap, reducing its value to 0.59 eV and inducing a magnetic moment of 3 $\mu_B$. This magnetization is triple that induced by another symmetry-breaking strategy, surface adsorption of hydrogen or fluorine atoms (1 $\mu_B$),[63] and matches the behavior of the bulk wurtzite GaN counterpart.[64] The defect states arise from three holes generated per gallium vacancy, enabling efficient electron capture and charge transfer critical for spintronic functionality. The spin-down character of both band edges positions this defect-engineered system as a high-efficiency spin filter for spin-up electrons. Introduction of $V_{Ga}$ defects is therefore shown as a versatile means for engineering magnetism and spin selectivity in g-GaN.

Figure S3 reveals the origin of magnetization in the $V_{Ga}$ system by showing the total and projected DOS of nitrogen atoms around the vacancy. The main contribution to the defect states comes from the p orbitals, especially $p_x$ and $p_z$ orbitals, of the first-nearest-neighbor (1NN) nitrogen atoms of the defect site. In contrast, the second-nearest-neighbor (2NN) atoms exhibit negligible involvement, which is a hallmark of sharply localized magnetization. This phenomenon is further corroborated by the spin density isosurfaces presented in Figure S4. Bader charge analysis also presents a clear contrast: on average, the 1NN nitrogen atoms only gain 2.11 electrons each, while the rest of the nitrogen atoms each gain 2.98 electrons and show a strong ionic character. This orbital-selective spin polarization underscores the critical role of lattice symmetry breaking at the vacancy site, confining magnetism to the immediate defect environment. Such atomic precision in spin localization will allow designing of magnetically active nanoregions within GaN, enabling targeted spintronic functionality without long-range magnetic order.

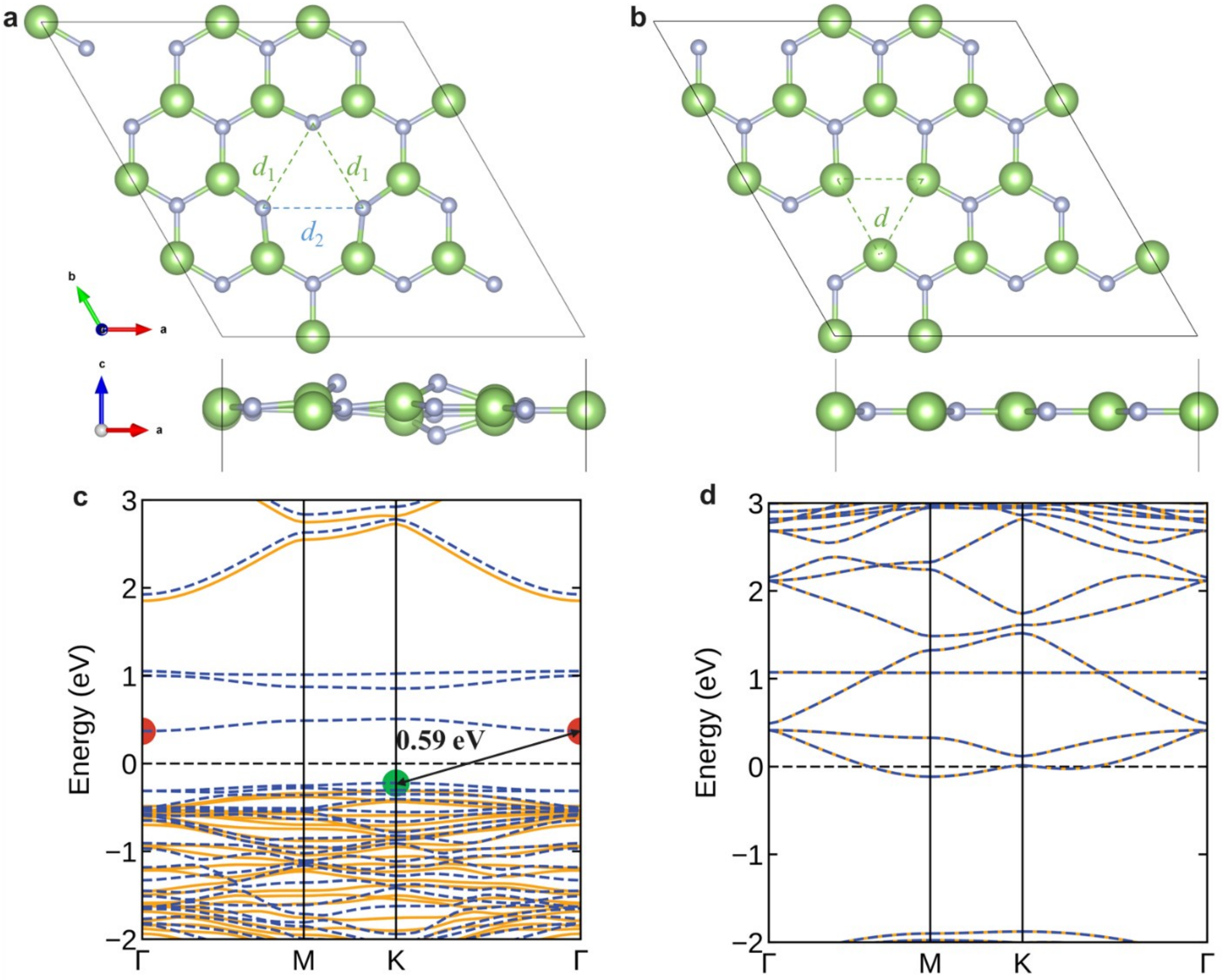

Figure 1. Properties of g-GaN with vacancy defects. Geometry of the fully relaxed structures: (a) $V_{Ga}$ (buckled) and (b) $V_N$ (planar) systems. The N–N and Ga–Ga interatomic distances around the vacancies are labeled accordingly. Electronic band structures: (c) $V_{Ga}$ system with an indirect band gap of 0.59 eV; (d) $V_N$ system with no band gap. Yellow solid line: spin-up bands; blue dashed line: spin-down bands. The zero energy is calibrated to be the Fermi level. The conduction band minimum and valence band maximum are labeled in red and green, respectively.

In contrast to the magnetic $V_{Ga}$ system, the $V_N$ defect induces a nonmagnetic state (Figure 1d). In addition to a flat defect state induced within the conduction band, two more defect states are formed and intersect the Fermi level. A similar phenomenon occurs in the case of bulk wurtzite GaN with the $V_N$ defect, where the Fermi level is also embedded in the conduction band.[65] Bader charge analysis shows that instead of losing all three valence electrons, the 1NN gallium atoms on average only lose 2.03 electrons each. This defect-driven metallization induces high charge carrier density at the Fermi level. Such modulation suggests $V_N$ defects could act as atomic-scale conductive pathways in g-GaN, enabling applications in low-resistance interconnects or plasmonic devices. The duality of the magnetically active $V_{Ga}$ system versus the metallized $V_N$ system showcases defect engineering as a powerful tool to tailor the electronic performance of g-GaN for divergent technological niches.

***Gas adsorption on pristine and defective g-GaN***

Many 2D materials demonstrate excellent gas sensing abilities. The selectivity of g-GaN has been previously demonstrated, with different induced magnetic moments, charge transfer behaviors, and material band gaps for common gases including NO, nitrogen dioxide ($NO_2$), carbon monoxide (CO), and ammonia ($NH_3$),[18] serving as its major advantage in gas sensing. The adsorption of gas molecules can also be used to modulate the properties of 2D materials. The gas adsorption behavior at the g-GaN surface and its interaction with material defects and electric fields are studied here in detail by using NO as an example. This gas exhibits a simple molecular structure, enabling more tractable modeling. As it is a toxic gas, the results obtained will be highly relevant for its sensing and capturing applications.

Four potential high-symmetry adsorption sites are considered, which are above the gallium atom, nitrogen atom, Ga–N bond, and center of the hexagon. The configuration with the lowest energy is the most stable and therefore corresponds to the most probable adsorption site. The adsorption does not cause any out-of-plane atomic displacement, *i.e.*, the monolayer remains planar. The most stable position of the NO gas molecule is above the Ga–N bond, where it is inclined at an angle of 34.0° with respect to the monolayer (Figure 2). The relatively high adsorption height $h$ (2.283 Å) and small magnitude of the adsorption energy $E_a$ (−0.325 eV) indicate physisorption. Note that the configuration of the NO molecule adsorbed directly above the nitrogen atom is also quite stable, with the adsorption energy being only 0.2 meV lower in magnitude.

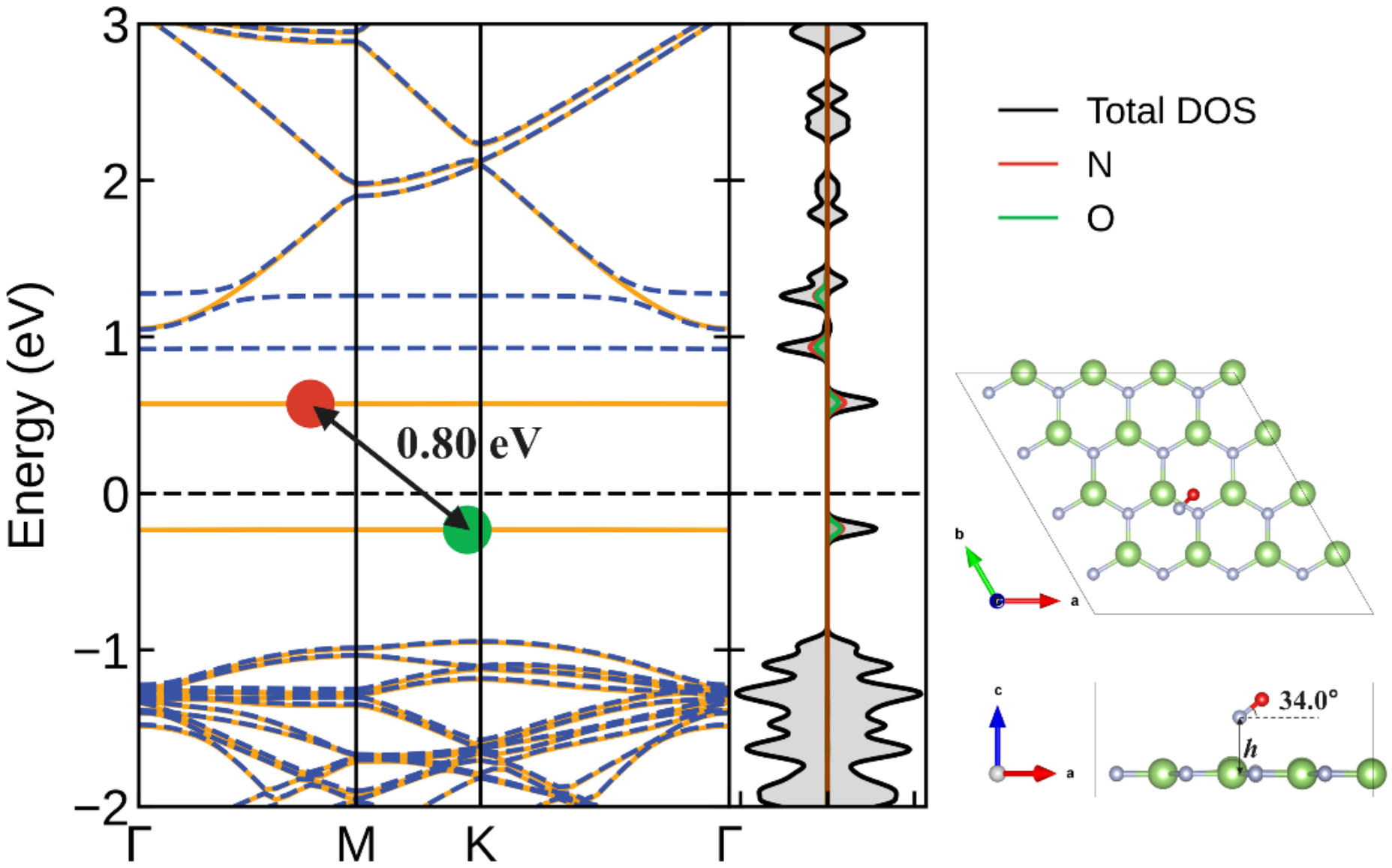

Figure 2. Spin-polarized electronic band structure of pristine g-GaN with adsorbed NO, showing a band gap of 0.80 eV (yellow solid line: spin-up states; blue dashed line: spin-down states). Corresponding total DOS and projected DOS of the adsorbed NO molecule (red: N; green: O). Geometry of the fully relaxed structure with the adsorption height *h* labeled.

Based on the transient state theory, the theoretical recovery time $\tau$ is estimated to be on the order of $10^{-7}$ s at 300 K using [66, 67]

$$\tau = \nu_0^{-1} \exp\left(-\frac{E_a}{kT}\right) \tag{1}$$

where $\nu_0$ is the assumed attempt frequency (~$10^{12}$ $s^{-1}$ for visible light),[67-70] $k$ is the Boltzmann constant ($8.617 \times 10^{-5}$ $eV \cdot K^{-1}$), and $T$ is the temperature (K). Compared to the theoretical values for other sensing materials [0.5 μs–16 s for $NO_2$ on carbon nanotubes,[67] 21 ms for NO on g-$BC_3$,[71] 205 μs for hydrogen ($H_2$) on palladium-doped $MoS_2$,[69] 4.7 ms for hydrogen sulfide on scandium-doped black phosphorene,[72] *etc.*], this result suggests the potential for rapid sensor recovery in g-GaN.

As shown in Figure 2, the adsorption of NO gas introduces three spin-polarized flat states within the band gap that significantly reduce the gap to 0.80 eV and impart a magnetic moment of 1 $\mu_B$. This change in the band gap $E_g$ increases the material conductivity $\sigma$ by 10 orders of magnitude at 300 K as determined by [73]

$$\sigma = AT^{3/2}\exp\left(-\frac{E_g}{2kT}\right) \tag{2}$$

where $A$ is a system-dependent constant of proportionality (electron $m^{-3}\cdot K^{-3/2}$), demonstrating the high sensitivity of the material to NO. The magnetization rivals the value induced by the above-mentioned strategy of surface adsorption of hydrogen or fluorine atoms but with the added advantage of reversible gas-responsive tuning. DOS analysis reveals strong hybridization between the orbitals of NO and GaN. This interplay demonstrates the potential of the material in gas sensing and on-demand spintronic switching, where the adsorbed molecule directly modulates both electronic and magnetic properties, unlocking opportunities for adaptive nanoelectronic systems.

Since vacancy defects are shown to not cause significant deformation to the crystal structure of g-GaN (Figure 1a and b), the defect systems could still provide a good base for adsorbing gas molecules. As shown in Table 1, defective g-GaN outperforms its pristine counterpart in NO adsorption. The vacancies drastically enhance the adsorption stability by reducing $h$ and increasing the magnitude of $E_a$. As shown in Figure 3a, direct vacancy occupation occurs in the $V_{Ga}$ system. With N–N bonds (1.404 Å) formed and the highly exothermic adsorption energy (−3.716 eV), chemisorption takes place, indicating irreversible NO capture. As the NO molecule experiences a higher diffusion barrier in the $V_{Ga}$ system (1.76 eV) than on pristine g-GaN (0.17 eV, Figure 4a and b), the former adsorption system is more stable and prevents NO escape. In contrast, although containing the N atom, the NO molecule does not occupy the nitrogen vacancy in g-GaN but becomes suspended 1.798 Å above the monolayer, and its N–O bond is perpendicular to the plane (Figure 3b). An energy barrier of only 0.99 eV is required for the NO molecule to diffuse away (Figure 4c). Weaker physisorption results. This remarkable divergence underscores $V_{Ga}$ defects as atomic-scale traps for NO gas molecules, leveraging vacancy-driven bond reconstruction to achieve high adsorption efficiency. Given the toxicity and prevalence of NO in industrial and urban emissions, this irreversible adsorption suggests the potential of $V_{Ga}$-engineered ultrathin GaN in air purification systems by offering a promising approach to neutralize hazardous gases in environmental and occupational settings.

Table 1. Magnetic, electronic, and NO gas sensing properties of pristine and defective g-GaN, obtained using the Perdew–Burke–Ernzerhof (PBE) functional.

| | **Pristine** | $V_{Ga}$ | $V_N$ |
|---|---|---|---|
| Magnetic moment ($\mu_B$) | 0 | 3 | 0 |
| Band gap (eV) | 1.94 | 0.59 | 0 |
| **NO adsorption** | | | |
| $h$ (Å) | 2.283 | 0.591 | 1.798 |
| $E_a$ (eV) | −0.325 | −3.716 | −1.360 |
| Diffusion barrier (eV) | 0.17 | 1.76 | 0.99 |
| Magnetic moment ($\mu_B$) | 1 | 0 | 2 |
| Band gap (eV) | 0.80 | 0.47 | 0.75 |

G-GaN exhibits a higher $E_a$ magnitude than other 2D materials commonly used for gas sensing, such as $MoS_2$ (−0.14 eV)[74] and graphene (−0.30 eV),[14] indicating its better gas capturing capability. Vacancy introduction as a measure to further improve the capturing is much more effective in g-GaN than in $MoS_2$ ($E_a$: −2.57 eV) and graphene ($E_a$: −3.04 eV). While the high $E_a$ magnitude of the $V_{Ga}$ system hinders gas desorption for device recovery ($\tau$ on the order of $10^{50}$ s), it is advantageous for applications where high sensitivity is preferred, and irreversible detection/capture is acceptable (*e.g.*, industrial leak alarms). For continuous monitoring, the weaker binding offered by the $V_N$ system or hybrid architectures may be preferable. Heat, ultraviolet light illumination, and voltage pulses are common strategies to provide the energy required for desorption. For example, upon heating to 300 °C, with $E_a = -1.360$ eV, the recovery time of the $V_N$ system can be significantly reduced to 0.91 s according to eq (1).

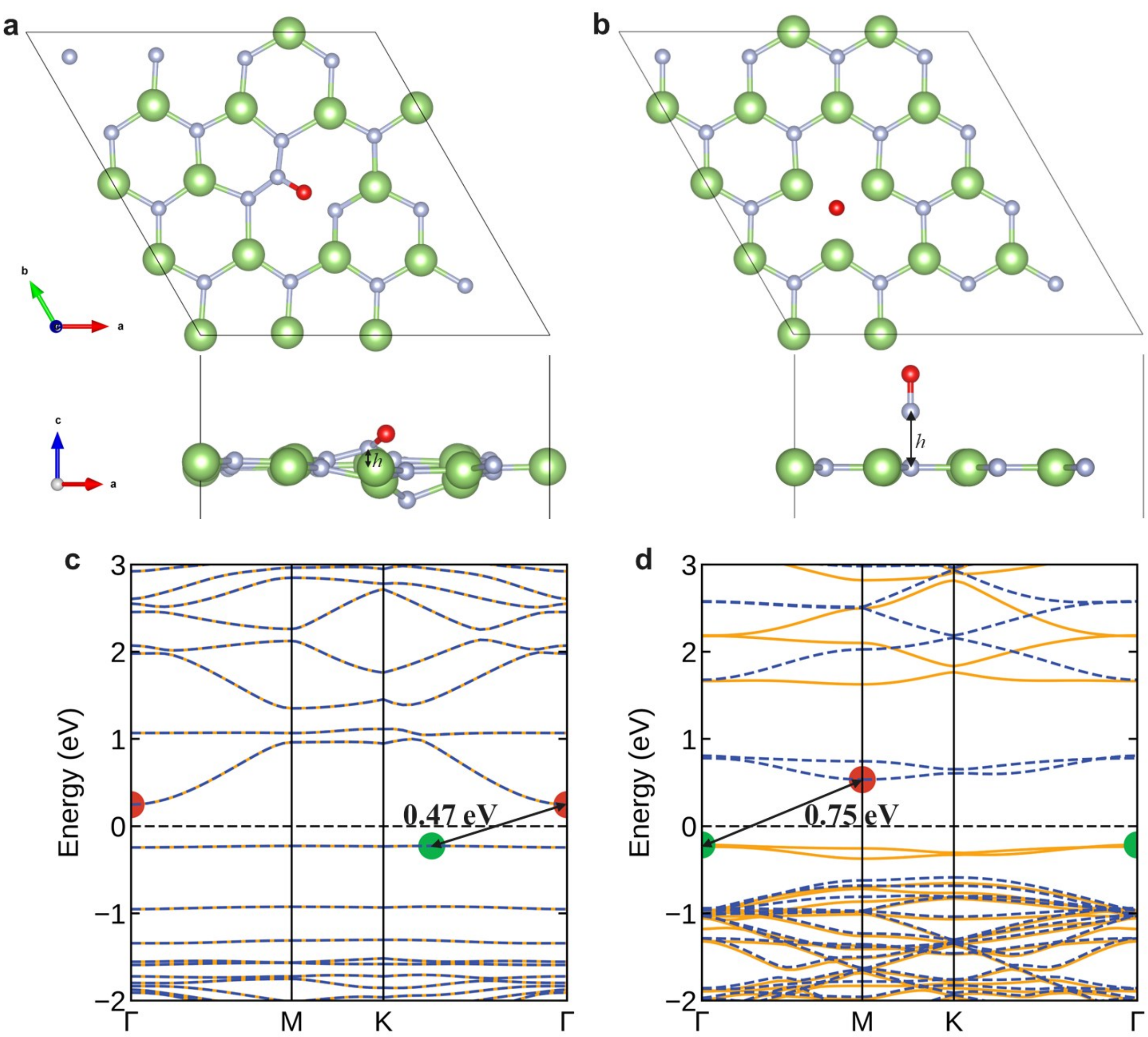

Figure 3. Adsorption of NO gas on g-GaN with vacancy defects. Geometry of the fully relaxed structures: (a) NO–$V_{Ga}$ and (b) NO–$V_N$ systems. Electronic band structures: (c) NO–$V_{Ga}$ and (d) NO–$V_N$ systems. Yellow solid line: spin-up states; blue dashed line: spin-down states.

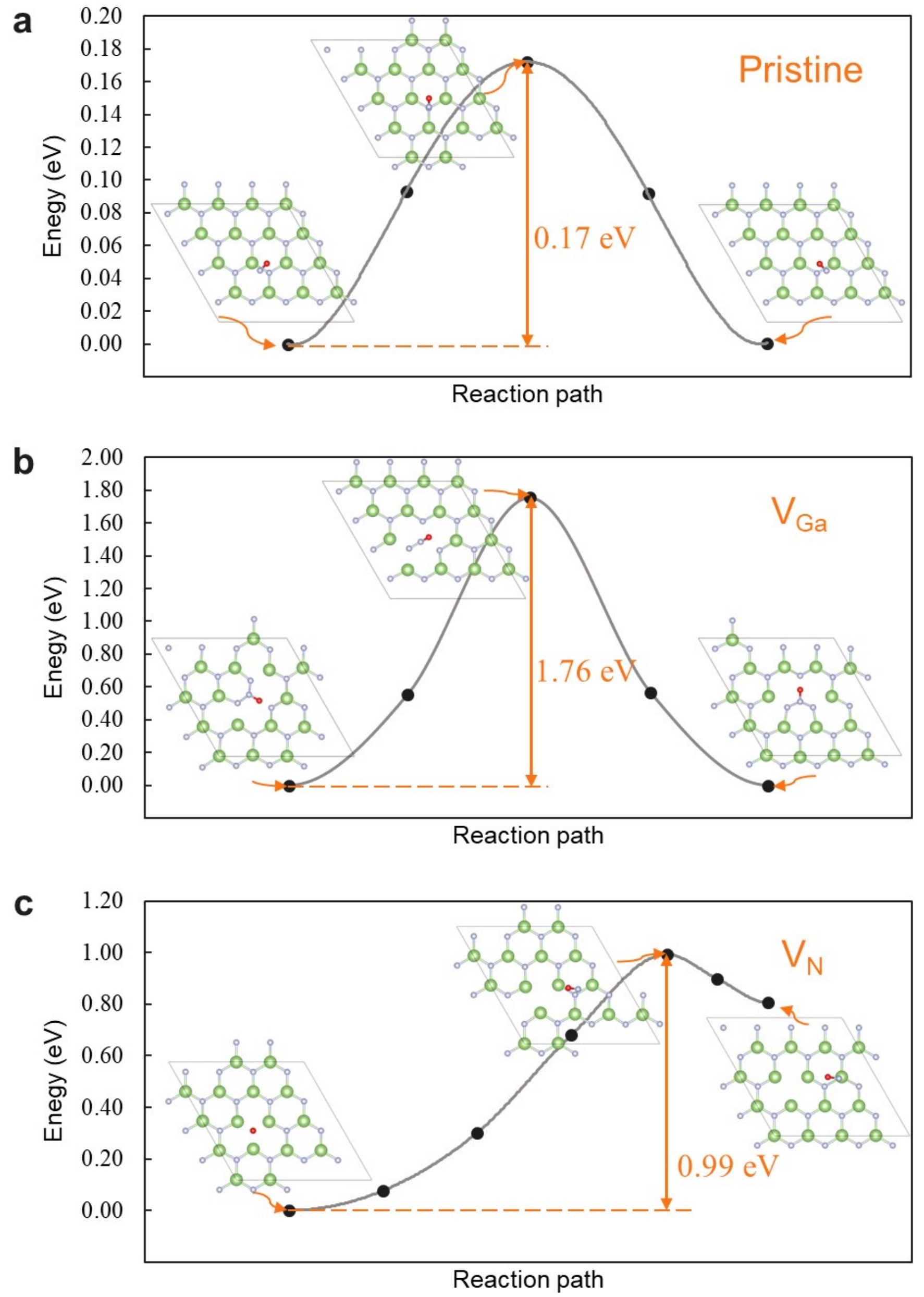

Figure 4. Diffusion path of a NO molecule (a) on pristine g-GaN, (b) in the $V_{Ga}$ system, and (c) in the $V_N$ system with the corresponding energy barriers calculated using the nudged elastic band method.

The adsorption of NO gas leads to defect-dependent electronic and magnetic responses in g-GaN (Figure 3c and d). Three flat valence bands emerge in the NO–$V_{Ga}$ system, reducing the band gap to 0.47 eV. The system becomes nonmagnetic. Conversely, NO adsorption on the $V_N$ system results in a magnetic moment of 2 $\mu_B$ *via* spin-polarized band splitting, with distinct spin-up (1.84 eV) and spin-down (1.12 eV) gaps. The completely different magnetization induced by NO adsorption on pristine g-GaN (1 $\mu_B$), $V_{Ga}$ (0 $\mu_B$), and $V_N$ (2 $\mu_B$) systems enables this gas to act as a magnetic probe for identifying vacancy types in g-GaN. The defective systems exhibit amplified magnetization changes upon NO exposure compared to pristine g-GaN, advocating their ability to produce stronger magnetic signals for gas detection. These computational insights provide a baseline for g-GaN in gas sensing and capturing; however, a complete assessment of its practical feasibility will require future investigations into charged-defect thermodynamics, the evaluation of sensing selectivity in multi-species environments, and experimental validation against established benchmark materials.

To explain the mechanism behind the above results, the charge density difference $\Delta\rho$ of the adsorption systems is calculated as

$$\Delta\rho = \rho_{\text{total}} - \rho_{\text{M}} - \rho_{\text{g-GaN}} \tag{3}$$

where $\rho_{\text{total}}$, $\rho_{\text{M}}$, and $\rho_{\text{g-GaN}}$ are the charge densities of the adsorption system, gas molecule, and g-GaN supercell, respectively. Figure 5 presents the comparison between the pristine and defective systems. The defective systems show more interaction with the gas molecule, especially in the case of $V_{Ga}$, where significant charge redistribution occurs not only around the gas molecule but also within the monolayer near the vacancy.

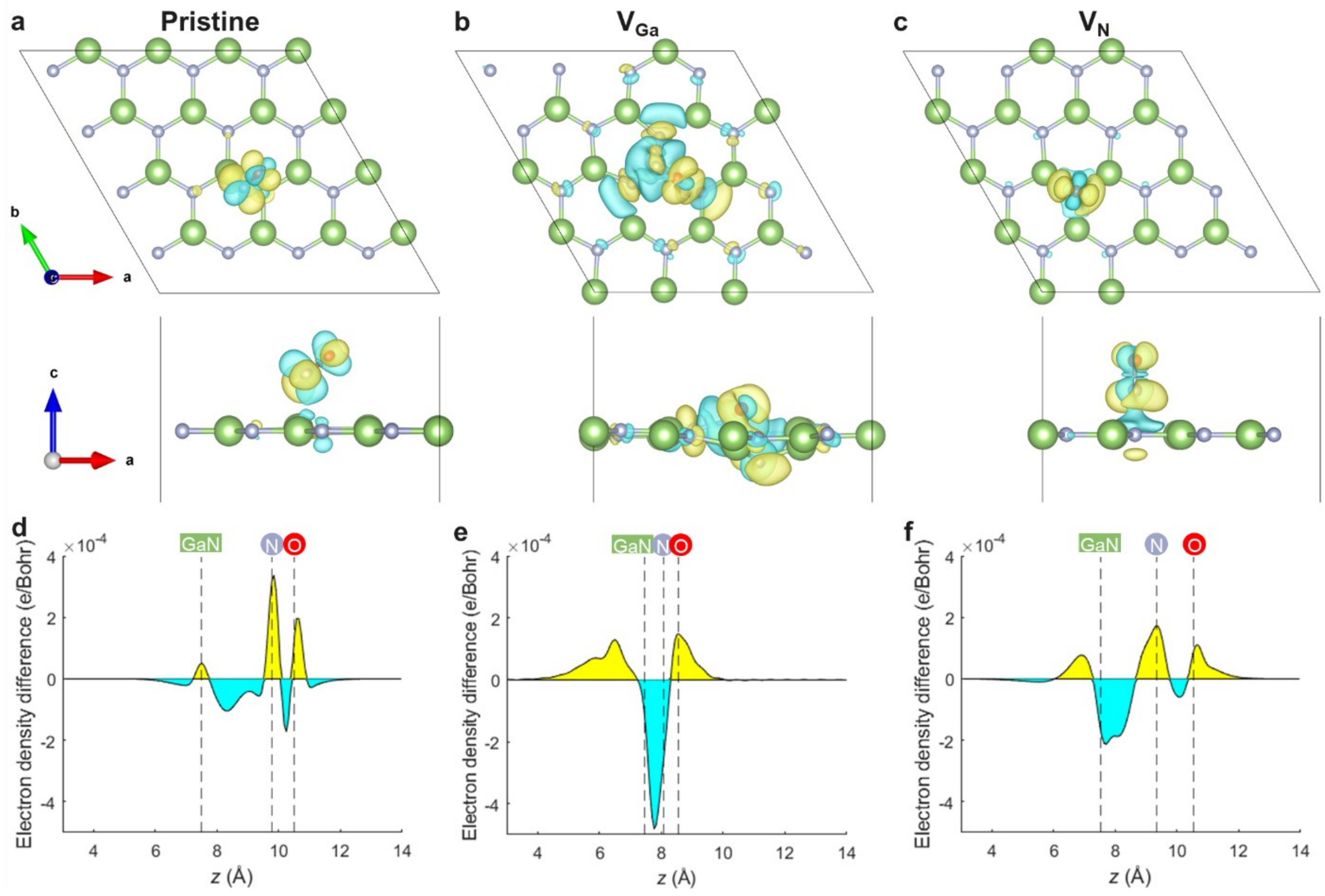

Figure 5. Charge density differences in NO gas adsorption systems. Three-dimensional isosurfaces (0.003 |e|·$Bohr^{-3}$) corresponding to NO adsorbed on (a) pristine g-GaN, (b) $V_{Ga}$, and (c) $V_N$ systems. Plane-averaged charge density difference along the normal (*z*) direction corresponding to NO adsorbed on (d) pristine g-GaN, (e) $V_{Ga}$, and (f) $V_N$ systems. The positions of the GaN monolayer as well as the N and O atoms of the NO molecule are labeled accordingly. Cyan: electron depletion; yellow: electron accumulation.

### *Impact of electric fields*

Building on the insights gained from the investigation into the effects of point vacancies and gas molecule adsorption on g-GaN, it is recognized that these methods, while effective for property modification, do not lend themselves to dynamic control. To bridge this gap, the application of electric fields emerges as a versatile tool and facilitates rapid modulation of properties. This modulation is usually achieved by manipulating the electronic structure, optical characteristics, phase behavior, and valley polarization of the 2D material.[75] For example, the application of a high electric field has been shown to close the band gap in various bilayer transition metal dichalcogenides, including $MoS_2$, $MoSe_2$, $MoTe_2$, and $WS_2$.[76] Likewise, a vertical electric field has been observed to significantly modulate the charge transfer between a range of gas molecules (*e.g.*, $H_2$, $O_2$, $H_2O$, $NH_3$, NO, $NO_2$, and CO) and monolayer $MoS_2$.[77] In view of the wide high-voltage applications of GaN and its promising capability in gas detection, it is both timely and pertinent to explore how an electric field can further refine the functionalities of g-GaN in electronic, magnetic, and gas sensing domains. This section therefore delves into the nuanced ways in which a vertical electric field can be harnessed to optimize the performance of g-GaN in these applications.

The electronic properties of pristine g-GaN exhibit high stability against vertical electric fields, maintaining a stable band gap up to a critical threshold $E_c$ of 1.0 $V \cdot Å^{-1}$ (Figure 6a). Beyond this point, an abrupt linear decrease in the band gap is observed. Semiconductor-to-metal transition occurs at around 1.5 $V \cdot Å^{-1}$ ($E_t$), which is attributed to the field-induced near-free electron gas.[47] This $E_t$ value represents the intrinsic limit of the perfect lattice considered here, thus establishing the theoretical upper bound for the electronic stability of the material. The semiconductor-to-metal transition is further evidenced by the plane-averaged potential shown in Figure S5. Oscillations above the monolayer start to occur at $E_t$ because of electron spill-out, which indicates the onset of field emission. Compared to bulk GaN that has an $E_t$ value of 0.03–0.05 $V \cdot Å^{-1}$,[78, 79] g-GaN exhibits at least a 30-fold improvement, allowing the material to withstand a much higher voltage before the onset of its metallic character. The dynamic stability of g-GaN under such intense electric fields is also verified with phonon calculations, which show no imaginary frequencies (Figure S6).

Because of the field emission phenomenon, where a metal surface emits current in an electric field on the order of 0.1 $V \cdot Å^{-1}$, it is challenging to generate high electric fields without forming an electric arc, especially in media with low dielectric strength like air. However, it is still possible to realize strong electric fields to engineer the band gap of g-GaN. One method is to start with a much weaker field generated by cathode and anode plates in vacuum, and between these, place an electrode with a sharp radius of curvature at the tip on the order of 1 μm or lower. With the curvature of the field near the tip, it can reach orders of magnitude higher than the surroundings.[80] One approach that bypasses the breakdown of gate dielectrics to produce intense electric fields is to suspend the 2D material between two volumes of ionic liquid with independently controlled potentials.[81] The field-tunable band gap modulation, paired with the ability to retain semiconducting behavior under intense biases, expands the applications of g-GaN for high-voltage nanoelectronics where conventional bulk semiconductors fail.

Upon the application of an electric field, additional states emerge in the conduction band (Figure 6b). As the electric field increases up to $E_c$, these states keep their energy well above the conduction band minimum, and the band gap does not change much. Since the shape of the conduction band edge is barely changing, the electron effective mass, which is inversely proportional to the band curvature, also remains relatively constant (Figure 6c). Beyond $E_c$, the additional states proceed to move toward the Fermi level, progressively displacing the original conduction band edge and driving a controlled band gap closure. This field-induced conduction band restructuring occurs without distorting the valence band, whose edge remains inert to electric perturbations. The decoupled response of dynamic conduction band modulation paired with a rigid valence band enables precise tuning of the electron transfer. The band gap collapse beyond $E_c$ offers a gateway to on-demand metallicity.

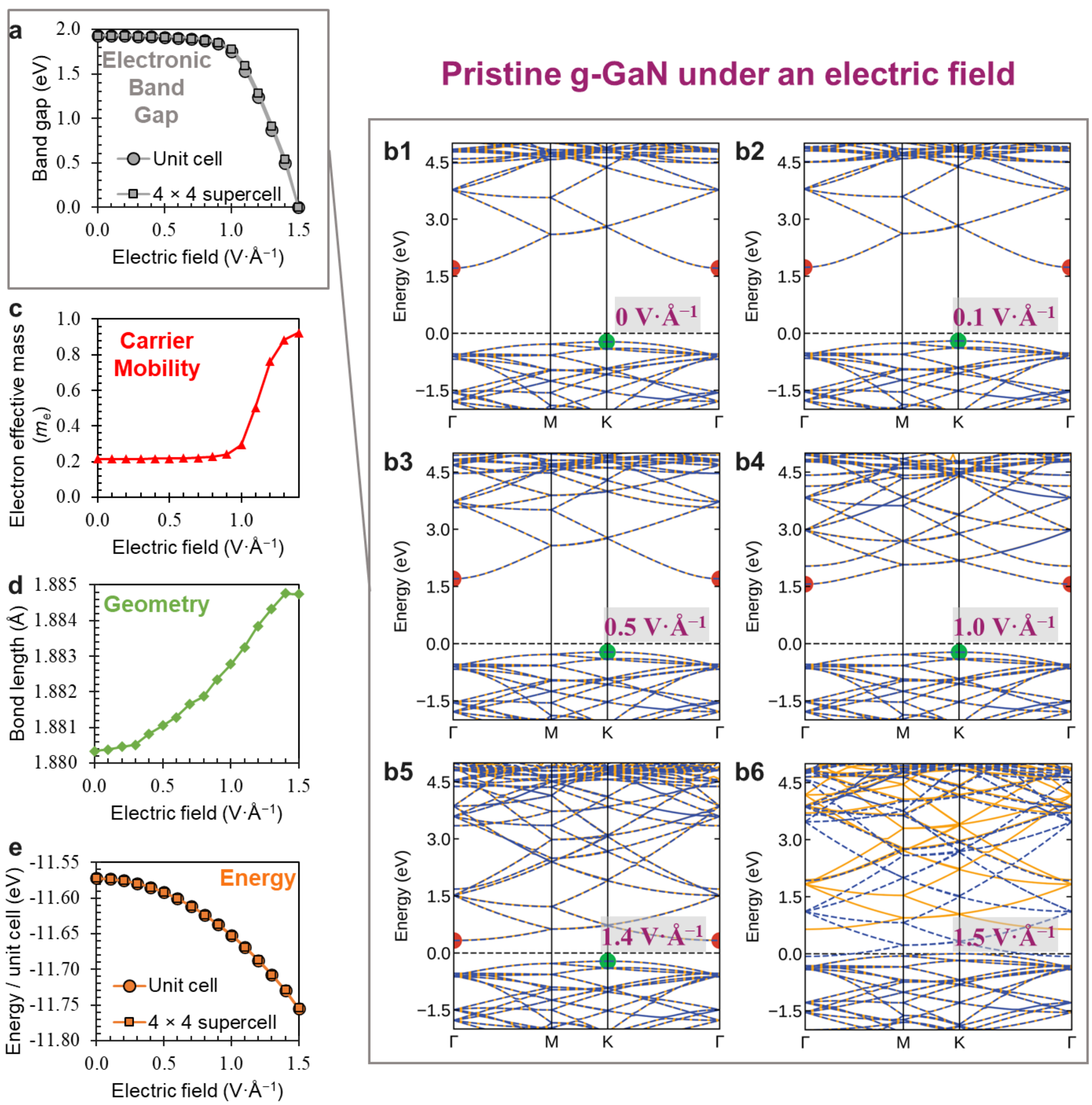

Figure 6. Changes in the properties of pristine g-GaN under a vertical electric field of up to 1.5 V·Å$^{-1}$. (a) Electronic band gap (unit cell and 4 × 4 supercell). (b) Electronic band structures of the 4 × 4 supercell under electric fields of (1) 0, (2) 0.1, (3) 0.5, (4) 1.0, (5) 1.4, and (6) 1.5 V·Å$^{-1}$. Yellow solid line: spin-up states; blue dashed line: spin-down states. (c) Electron effective mass (4 × 4 supercell). (d) Ga–N bond length. (e) Energy (unit cell and 4 × 4 supercell). The results obtained by using both the unit cell and 4 × 4 supercell are consistent with one another.

As shown in Figure 6d, the Ga–N bond length increases with the electric field, similar to the effect of applying tensile strain. Extending the Ga–N bond reduces the interaction between the neighboring gallium and nitrogen atoms as well as the overlap of the adjacent atomic orbitals, which has been shown to decrease the indirect band gap between the K and Γ sites.[82] Figure 6e shows that an electric field decreases the system energy in a parabolic manner, signaling enhanced stability under extreme electric loads.

The influence of strain on the band gap evolution under an external electric field was investigated to assess the performance under operational conditions (Figure S7). While the band gap of strained g-GaN follows the trend of the pristine structure, the different nature of the strain leads to varying impacts other than changing the initial band gap values. In-plane compression causes early occurrence of $E_c$ and $E_t$, while tension causes the band gap to converge with that of the pristine structure above 1.2 $V \cdot Å^{-1}$. These results indicate that compressive strain facilitates low-field metallization, whereas tension is preferable for applications requiring stable electronic behavior.

The performance of GaN is often compared with that of another wide-band-gap semiconductor, silicon carbide (SiC), for power electronics applications. While SiC reacts with oxygen to form a protective silica ($SiO_2$) layer on the surface that enhances its stability, 2D GaN also exhibits high thermal stability up to 800 °C before oxidation.[28] SiC has a stable 2D honeycomb form, g-SiC. The magnetic and electronic properties of g-GaN and g-SiC monolayers in pristine and defective forms are compared in Table 2. Both materials are nonmagnetic in their pristine forms. While the band gap of g-GaN is clearly indirect, it is difficult to classify the band gap of g-SiC as indirect or direct because of its flat lowest conduction band.[85] Upon application of a vertical electric field, both materials experience a semiconductor-to-metal transition. The transition is however much delayed in the case of g-GaN, occurring at an electric field three times higher, making the finite band gap more electric field–resistant. In both systems, vacancy defects can induce magnetism. The g-GaN monolayer experiences more drastic band gap changes than g-SiC after the introduction of vacancies, indicating its potential to be functionalized with novel electronic properties.

Table 2. Property comparison between g-GaN and g-SiC monolayers.

| | **G-GaN** | | **G-SiC** | |
|---|---|---|---|---|
| Magnetization ($\mu_B$) | 0 | | 0 | |
| Band gap (eV) | 1.94 (indirect) | | 2.53–2.58 (flat band) [83-85] | |
| Effect of electric fields | Band gap vanishes at 1.5 V·Å$^{-1*}$ | | Band gap vanishes at >0.5 V·Å$^{-1*}$ [85] | |
| Vacancy | **$V_{Ga}$** | **$V_N$** | **$V_{Si}$** | **$V_C$** |
| Magnetization ($\mu_B$) | 3 | 0 | 4 [83] | 0 [83] |
| Band gap (eV) | 0.59 | 0 | 1.32 [86] | 2.54 [84] |

*Obtained based on the PBE functional.

To explore the full potential of vacancy defects, we further investigated the responses of defective g-GaN to electric fields. In the $V_{Ga}$ system, the changes of the band gap with the electric field can be divided into three stages (I, II, and III) as shown in Figure 7a, and the band structures at selected electric fields are presented in Figure 7b. Before the electric field increases to 0.7 V·Å$^{-1}$, the band gap remains steady at around 0.59 eV (stage I). As illustrated in Figure 7b2, there is an occupied defect state dominating the edge of the valence band. The valence band maximum, as indicated by the red dot, therefore lies along the defect band. In stage II, the band gap steps up slightly to 0.65 eV, which is absent in the case of pristine g-GaN, and starts to decrease slowly. The small increase is attributed to the defect state being embedded in the valence band as shown in Figure 7b3. Beyond an electric field of 1.4 V·Å$^{-1}$, the band gap quickly drops and becomes zero at around 1.6 V·Å$^{-1}$ (stage III). Given that the thickness of the $V_{Ga}$ monolayer under this electric field is ~1.15 Å, the electric field corresponds to a high gate bias or voltage drop of ~1.8 V. As shown in Figure 7b4–6, further increase in the electric field causes the conduction band to quickly shift down with respect to the Fermi level, thereby embedding the unoccupied defect states and eventually closing the band gap.

Introducing the $V_{Ga}$ defect does not significantly alter the electric field at which the semiconductor-to-metal transition occurs; however, $E_c$ significantly increases. Since the energy of the defect states dominating the band edges is not sensitive to changes in the electric field, before other higher-energy states shift down and become the conduction band edge, the band gap remains relatively constant. This robust performance is therefore uniquely enabled by the defect, demonstrating the cooperative mechanism of vacancy defect–stabilized band gap under intense electric fields. Together with the steady carrier mobility, a high $E_c$ value can ensure the stable performance of the material even under high alternating-current voltage, a critical advantage for power electronics. Notably, even at metallization-onset fields (1.6 $V \cdot Å^{-1}$), a finite spin-up band gap of 0.82 eV persists. This phenomenon signals spin-selective semi-metallicity, which can be exploited for spin-filtering devices.

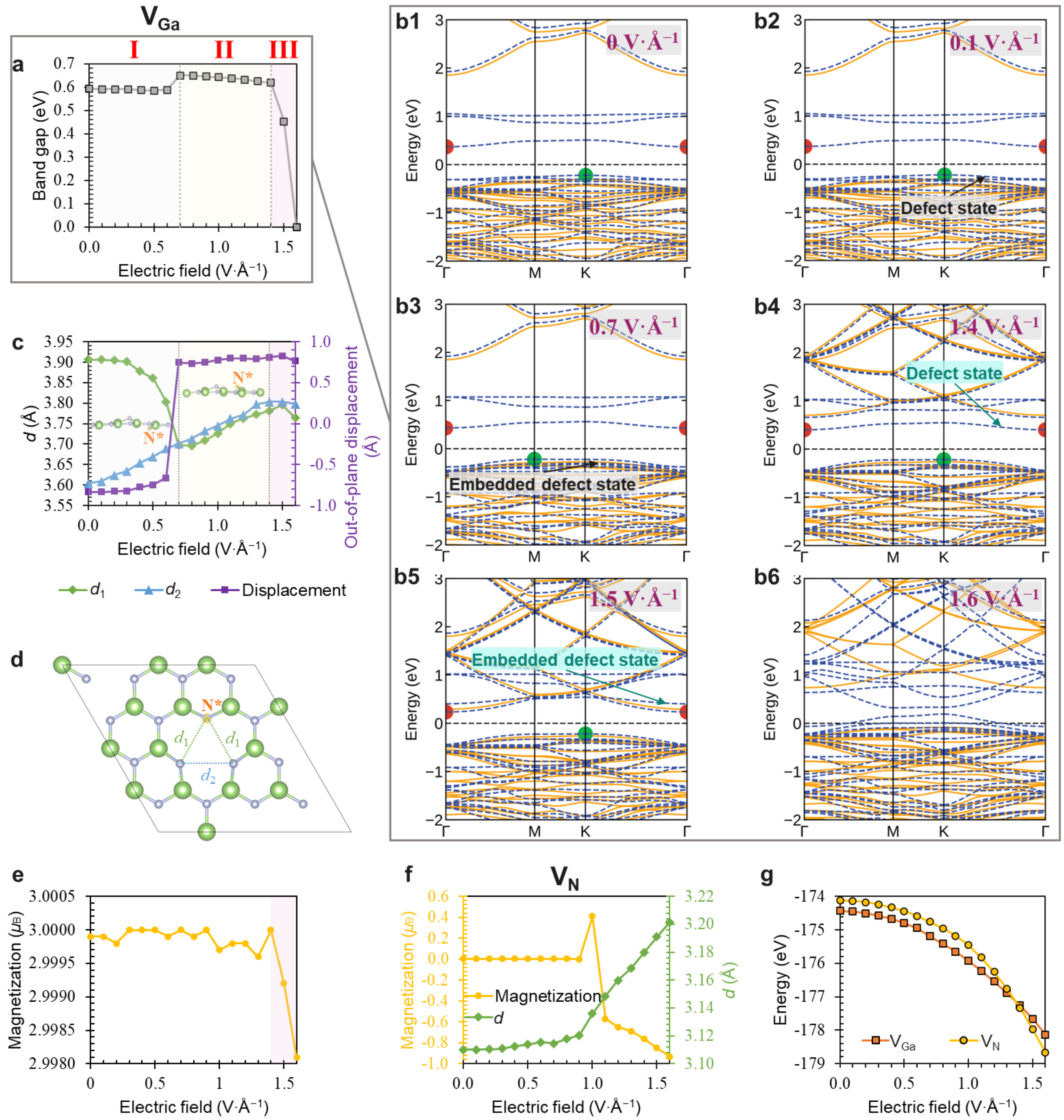

Figure 7. Changes in the properties of defective g-GaN under a vertical electric field of up to 1.6 V·Å$^{-1}$. (a) Electronic band gap of the $V_{Ga}$ system. (b) Electronic band structures of the $V_{Ga}$ system under electric fields of (1) 0, (2) 0.1, (3) 0.7 (4) 1.4, (5) 1.5, and (6) 1.6 V·Å$^{-1}$. Yellow solid line: spin-up states; blue dashed line: spin-down states. (c) Interatomic distances $d_1$ (green diamond) and $d_2$ (blue triangle) between the nearest nitrogen neighbors of $V_{Ga}$ and out-of-plane displacement of N* (purple square). A negative value indicates that N* is located below the GaN monolayer. (d) Top-view schematic indicating the N* atom and interatomic distances $d_1$ and $d_2$. (e) Magnetization in the $V_{Ga}$ system. (f) Magnetization (yellow circle) in the $V_N$ system and interatomic distance $d$ (green diamond) between the nearest gallium neighbors of $V_N$. (g) Energy of the $V_{Ga}$ (orange square) and $V_N$ (yellow circle) systems.

Since it has been shown that 1NN nitrogen atoms are the main contributors to the defect states (Figures S3 and S4), these atoms, including their relative positions, and the resulting charge redistribution should be investigated to determine the reason behind the changes in the band structure under electric fields. As shown in Figure 7c and d, as one of the 1NN atoms (N*) is displaced to the opposite side of the monolayer relative to the other two with no electric fields, its distance from these two atoms $d_1$ is larger than the distance between these two atoms themselves $d_2$. As the electric field increases in stage I, N* is pulled back closer toward the monolayer as indicated by the decreasing magnitude of its out-of-plane displacement. The interatomic distance $d_1$ decreases while $d_2$ increases until both values are about the same, which is when all three 1NN atoms lie on the same side of the monolayer. In stage II, as the out-of-plane displacement increases in general, both $d_1$ and $d_2$ increase with the electric field, mirroring the tensile effect observed in the pristine case. In stage III, these values eventually saturate and even start to decrease slightly, leading to vanishing of the band gap. The fluctuations in the interatomic distances and the out-of-plane displacement of N* correspond well with the nonmonotonic changes in the band gap. This correlation underscores the critical influence of 1NN atoms on the electronic properties of the $V_{Ga}$ system. These atomic-scale insights into defect-mediated structure–property coupling provide a blueprint for defect engineering, where targeted atomic displacements could enable on-demand electronic phase switching.

The movement of N* and the resulting rearrangement of the states also affect the vacancy-induced magnetism. Figure 7e shows the highly stable magnetization as the electric field increases, which only starts to reduce slightly beyond $E_c$. While magnetism in the pristine and $V_{Ga}$ systems is not affected much by the electric field, it can be altered significantly in the case of the $V_N$ system. Below an electric field of 1.0 $V \cdot Å^{-1}$, the system remains nonmagnetic, while the nearest gallium neighbors of the vacancy slowly separate as indicated by the increase in $d$ (Figure 7f). Then, magnetization occurs with a positive value first, indicating that the magnetic moment is initially parallel to the electron spin. It becomes negative thereafter, and the magnitude increases. Meanwhile, $d$ surges abruptly. The onset of magnetization in the system can be attributed to the increase in $d$ beyond a certain threshold, which overcomes the electrostatic attraction between gallium atoms around the vacancy, thereby releasing free electrons. The spin-flip transition could be due to a metastable state, where liberated electrons transiently occupy spin-up states before relaxing into spin-down configurations. The ability to toggle magnetism with electric fields introduces the potential for information to be not only stored but also erased and rewritten in a non-volatile manner for memory applications. By coupling structural distortion to spin polarization, defect-engineered g-GaN offers atomic-scale control over spin states for next-generation computing architectures.

Figure 7g presents the energy changes of the defective systems in response to the electric field. Both systems exhibit a decrease in energy as the electric field increases, consistent with the trend observed for pristine g-GaN.

The adsorption of gas molecules on pristine and defective g-GaN can also be modulated by vertical electric fields. With an increasing electric field up to the intrinsic limit of stability, more charge transfer occurs between g-GaN and the adsorbed NO molecule (Figure S8). As shown in Figure 8a, increasing the electric field causes the magnitude of $E_a$ in the pristine system to increase, making the adsorption more stable. Before the electric field increases to 1.0 $V \cdot Å^{-1}$, the magnitude of $E_a$ decreases in the NO–$V_{Ga}$ system but increases in the NO–$V_N$ system. Beyond this electric field, the opposite trend occurs for both defective systems. These changes present the potential to dynamically control the sensor recovery and gas capture rate through simply varying the applied voltage. Figure 8b shows that $h$ becomes less correlated with $E_a$ upon application of the electric field.

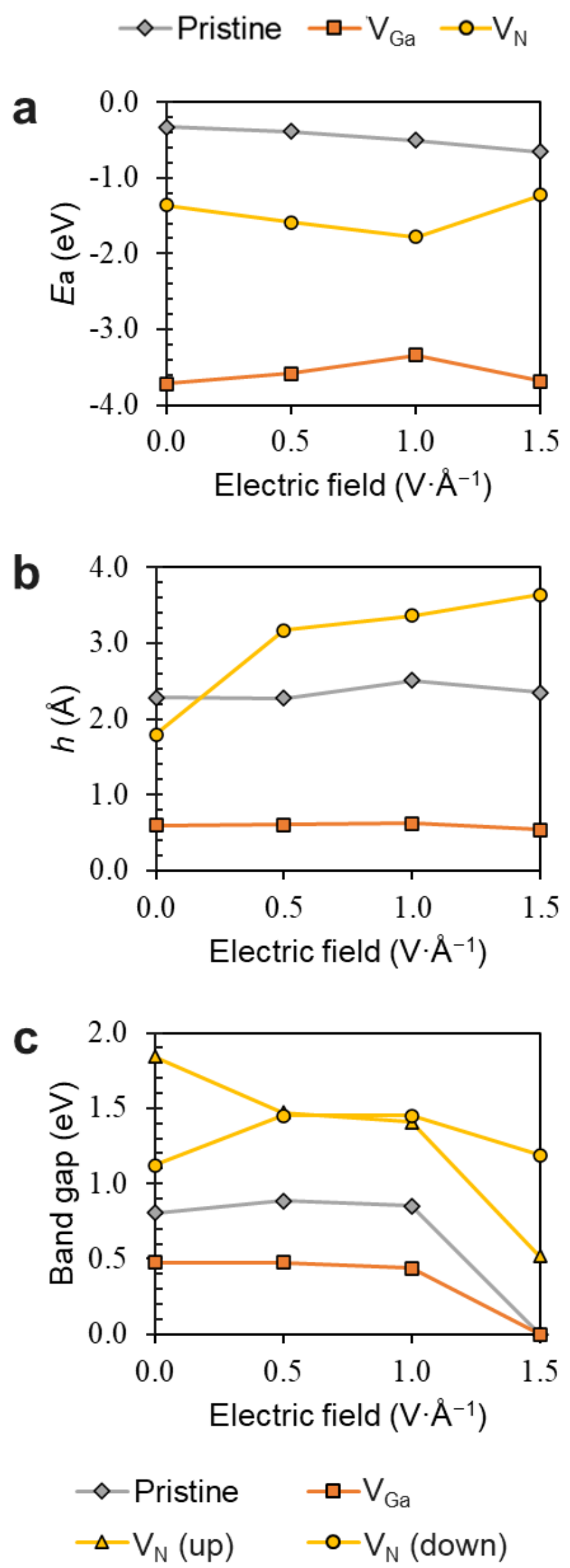

Figure 8. Changes in the properties of NO gas adsorption systems under a vertical electric field of up to 1.5 V·Å$^{-1}$. (a) Adsorption energy $E_a$. (b) Adsorption height $h$. (c) Band gap.

As depicted in Figure 8c, the band gap of g-GaN with an adsorbed NO molecule behaves differently under an increasing electric field compared to the pristine material; it does not simply decrease monotonically. Initially, there is a slight increase in the band gap, but as the electric field strength continues rising, the gap eventually drops to zero, signaling a semiconductor-to-metal transition. The band gap of the NO–$V_{Ga}$ adsorption system remains relatively constant at first, although it then vanishes as the electric field increases to 1.5 V·Å$^{-1}$. The spin-up and spin-down band gaps of the NO–$V_N$ adsorption system converge toward one another to around the same value as the electric field increases to 1.0 V·Å$^{-1}$. Beyond this point, both values decrease, but the spin-up band gap reaches a lower value at an electric field of 1.5 V·Å$^{-1}$. In the NO gas environment, while both the pristine and $V_{Ga}$ systems experience a semiconductor-to-metal transition eventually, the $V_N$ system maintains its semiconducting property even at high electric fields. The robust electronic behavior being not affected by the environmental gas would be highly relevant to maintaining the device performance under harsh environments with intense electric fields and the toxic NO gas.

With the adsorbed NO molecule, electric fields can no longer trigger changes in the magnetic moment of the $V_N$ system. As shown in Figure S9, electrons, which are released from Ga–Ga metallic bonds formed near the vacancy, are transferred to the electronegative N and O atoms in the molecule. There are thus no free electrons that can induce changes in magnetism.

***Targeted checks using the hybrid functional***

The Perdew–Burke–Ernzerhof (PBE) functional employed in this research is formulated within generalized gradient approximation (GGA) and offers an optimal balance between accuracy and computational cost. However, while the self-interaction error (SIE) in GGA is considerably smaller than in local density approximation,[87] such semilocal functionals still inherently suffer from this error, which can lead to the underestimation of band gaps, over-stabilization of delocalized electronic states,[88] and overbinding of simple adsorbates to solid surfaces.[89] To ensure our physical conclusions are robust against these systematic errors, targeted validation checks have been performed using the more computationally expensive Heyd–Scuseria–Ernzerhof functional HSE06.[90] By incorporating Hartree–Fock exchange at short range, HSE calculations mitigate the SIE, confirming that the limitations of the semilocal approximation do not compromise the central findings of this research regarding adsorption trends and the metallization threshold.

Since the $V_{Ga}$ system exhibits the strongest interaction with the NO molecule in our PBE analysis, it serves as the critical representative case for validating adsorption energetics and charge transfer. Our HSE06 calculations yield an adsorption energy of −2.56 eV for this system. While this magnitude is smaller than the PBE prediction (consistent with the correction of overbinding), it remains significantly larger than the values for the pristine surface or $V_N$ cases. This result confirms that the relative stability hierarchy is preserved, ensuring that our central claim of $V_{Ga}$ acting as a thermodynamic trap for NO is robust against the functional choice. Additionally, the significant charge transfer between the $V_{Ga}$ defect and the NO molecule is not only sustained but becomes more pronounced under the HSE functional (Figure S10), reflecting the superior ability of this functional to describe electron localization in defect states.[91]

The metallization behavior and magnetic property under electric fields were further validated. For the pristine structure, since Figure 6a demonstrates that unit cell and supercell calculations yield consistent trends, HSE validation for the pristine structure was performed on the unit cell to minimize computational cost. While a finite band gap of 1.05 eV persists at 1.5 $V\cdot Å^{-1}$ in the HSE results, semiconductor-to-metal transition occurs rapidly upon increasing the field to 1.6 $V\cdot Å^{-1}$. This result represents only a minor shift from the $E_t$ value predicted by the PBE functional. For the $V_{Ga}$ system (4 × 4 supercell), HSE checks confirm an identical $E_t$ value. These consistent results demonstrate that although the PBE functional underestimates the absolute band gap magnitude, it accurately captures the general physical trends and the critical field effects driving the semiconductor-to-metal transition.

While HSE calculations predict the same magnetic moments for pristine and $V_{Ga}$ systems under electric fields, they reveal that the $V_N$ system is actually magnetic with a magnetic moment of 1 $\mu_B$. This discrepancy indicates that the magnetic switching behavior observed in PBE calculations is an artifact of the tendency of the functional to over-delocalize electrons in the lattice. However, as the electric field increases, the PBE result converges to the HSE result, implying that the electric field physically distorts the lattice (by increasing $d$) enough to overcome the delocalization error. Thus, the PBE calculations at high fields are reliable and consistent with results obtained by hybrid functionals.

While hybrid functionals do not always yield significant improvements in ground-state structural properties and energetics compared to the PBE functional,[92] the HSE analysis presented here serves as a critical benchmark. The convergence of results between these two methods reinforces the reliability of the theoretical model, providing a solid predictive framework to guide future experimental validation.

## Conclusions

The first-principles approach is adopted to systematically reveal the synergy between realistic environmental and operational conditions, including defects, strain, environmental gases, and external electric fields in the case of a promising ultrathin material, g-GaN. The defect states introduced by point vacancies are shown to enhance the functional versatility of g-GaN; specifically, $V_{Ga}$ is predicted to improve electronic stability under varying electric fields. In-plane strain typically induced during fabrication or operation endows easier vacancy defect formation, and compressive strain facilitates metallization at significantly lower electric fields. We also explore the potential for gas sensing and capture using NO as a prototype, demonstrating that the adsorption interaction is defect-modulated and can be further amplified by an applied bias. This behavior demonstrates a potential strategy for both defect detection and precise property tuning *via* the concurrent effects of electric fields. While these findings establish the theoretical upper-bound behavior for g-GaN within the adopted DFT framework, we acknowledge that a full assessment of operational relevance requires further in-depth treatment of charged-defect thermodynamics and sensing selectivity. Ultimately, the transition to device-level applications will remain subject to experimental validation of these predicted stability and sensing limits.

Compared to prior studies that explored field-tunable properties in idealized GaN systems or defects alone, this work provides further mechanistic insight through investigating synergistic effects in non-ideal, application-relevant environments. By establishing a comprehensive framework to examine the geometry, energetics, band structure, DOS, magnetic moment, and charge transfer of the relevant material systems, we reveal the cooperative mechanisms such as the vacancy-stabilized band gap. We also demonstrate the outstanding operational robustness of this 2D phase of the wide-band-gap material under the concurrent impact of defects and electric fields, which is its previously unexplored strength for achieving device viability. Additionally, this work offers a pathway to exploit defects and fields constructively, marking a paradigm shift from traditional defect-mitigation strategies, and the atomistic insight derived may also be applicable to a broader material spectrum.

This systematic first-principles research focuses on the neutral limit to isolate the intrinsic magnetic coupling and adsorption physics, free from the background charge compensation effects. The established framework and derived insights provide a solid foundation for quantitatively exploring the impact of charged defects, defect clustering, and their interplay with environmental conditions, including electric and magnetic fields, different gaseous environments, strain, and temperature, to realize the full potential of g-GaN under even more realistic operating settings. A greater strain range can also be incorporated to investigate extreme cases and determine how the elastic limits can be impacted by various factors. These collective efforts will enable predictive design beyond idealized models to drive innovation and meet the growing demands of the dynamic fields of 2D materials and semiconductor technology.

**Methods**

While hybrid functionals may provide superior electronic accuracy, the computational cost precludes their use for the extensive diffusion and field-dependent screening performed here. Consequently, we employ the GGA functional method with the PBE formulation [93] for structural and kinetic predictions, validated by targeted HSE06 [90] checks which confirm that the central findings of this work remain robust. The DFT-D2 method by Grimme [94] is adopted to correct for van der Waals interaction. Standard Ga ($4s^2$ $4p^1$) and N ($2s^2$ $2p^3$) pseudopotentials are used. A plane wave basis set with a kinetic energy cutoff of 520 eV is selected. Periodic boundary conditions are applied in the in-plane (*x* and *y*) directions. With a lattice constant of 15 Å in the vertical (*z*) direction, free boundary conditions are enforced. Until the electric field increases to 1.4 $V·Å^{-1}$, a smooth linear slope is observed away from the monolayer in the plane-averaged potential (Figure S5), indicating that there is enough vacuum separation to prevent interaction between periodic images of the monolayer structure. A linear dipole correction is also employed to correct errors introduced by the periodic boundary conditions under electric fields. To model the vacancies and gas adsorption, a 4 × 4 × 1 supercell of g-GaN is constructed from its fully relaxed unit cell. The same supercell size was adopted in previous studies that investigated the effects of point defects and composition in hexagonal group-III nitride monolayers, including g-GaN,[39] and the adsorption of gas molecules like CO, $NH_3$, NO, and $NO_2$ on g-GaN [18] and InSe monolayers.[17] A k point convergence test was conducted, which determined the use of $14 \times 14 \times 1$ and $4 \times 4 \times 1$ Γ-centered k point grids for the unit cell and supercell, respectively. The structures are fully relaxed until the maximum Hellmann–Feynman force per atom is below 0.01 $eV \cdot Å^{-1}$, with the energy convergence criterion being $10^{-5}$ eV. Before incorporating the NO gas molecule into the structures to study its adsorption, the molecule is first fully relaxed with a single k point at the Γ point. The optimized bond length of the NO molecule is 1.169 Å.

Application of electric fields has been previously demonstrated to have negligible impact on the lattice constant of g-GaN.[95] In our trial calculation, only 0.2% change is observed in the lattice constant of pristine g-GaN under the highest electric field of 1.5 $V \cdot Å^{-1}$. Consequently, it is not optimized for each electric field. However, full relaxation of atomic positions is still performed to capture the induced geometry changes.

For calculation of electronic properties, the criterion for energy convergence is tightened to $10^{-6}$ eV. The DOS and band structure are calculated for each optimized material system. The high-symmetry path Γ–M–K–Γ within the Brillouin zone for the hexagonal lattice is adopted to calculate the band structure with 20 k points sampled along each segment of the path (Figure S1b). Since spin–orbit coupling has been previously reported to cause splitting of the degenerate bands by a negligible amount,[46] it is not considered here.

Several tools are adopted to post-process the results. VASPKIT is used to extract the spin density.[96] Bader Charge Analysis Code of version 1.05 is employed to quantitatively obtain the charge transfer information.[97-99] The DOS results of individual atoms are post-processed based on Radin's MATLAB codes.[100] Visualizations of the structures are done with the software Visualization for Electronic and STructural Analysis (VESTA).[101]

To determine whether the defective systems are energetically favorable, the formation energy $E_{\mathrm{form}}$ is calculated as

$$E_{\mathrm{form}}[\mathrm{V_{Ga}}] = E[\mathrm{V_{Ga}}] - E[\text{g-GaN}] + \mu_{\mathrm{Ga}} \tag{4}$$

$$E_{\mathrm{form}}[\mathrm{V_{N}}] = E[\mathrm{V_{N}}] - E[\text{g-GaN}] + \mu_{\mathrm{N}} \tag{5}$$

where $E[\mathrm{V_{Ga}}]$ and $E[\mathrm{V_N}]$ are the total energy of the $\mathrm{V_{Ga}}$ and $\mathrm{V_N}$ systems, respectively; $E[\text{g-GaN}]$ is the total energy of the g-GaN supercell; $\mu_{\mathrm{Ga}}$ and $\mu_{\mathrm{N}}$ are the chemical potentials of Ga and N atoms, obtained as the energy per atom in bulk Ga and a $N_2$ molecule, respectively.

To assess the stability of the adsorption of the NO gas molecule, the adsorption energy $E_{\mathrm{a}}$ is computed as

$$E_{\mathrm{a}} = E_{\mathrm{total}} - E_{\mathrm{M}} - E_{\text{g-GaN}} \tag{6}$$

where $E_{\mathrm{total}}$, $E_{\mathrm{M}}$, and $E_{\text{g-GaN}}$ are the energy of the system with the adsorbed gas molecule, isolated NO gas molecule, and g-GaN supercell, respectively. When an electric field is applied, $E_{\mathrm{a}}$ is computed with the energy obtained for all the structures under the same electric field. A higher magnitude of $E_{\mathrm{a}}$ indicates a more energetically favorable adsorption configuration.

Phonon calculations were performed to verify the dynamic stability of the structures under selected electric fields based on the finite-difference method with the assistance of Phonopy [102, 103] (2.38.1). The energy convergence criterion is tightened to $10^{-8}$ eV to obtain more precise results for fully relaxed structures. $5 \times 5 \times 1$ supercells with displacements are constructed from the unit cell fully considering crystal symmetry. Non-analytical term correction was performed to obtain information about the Born effective charge and dielectric constant. Phonon frequencies and eigenvectors are calculated from the dynamical matrices with $80 \times 80 \times 1$ q points.

## Acknowledgements

This work was supported by the Economic Development Board - Singapore and Infineon Technologies Asia Pacific Pte. Ltd. through the Industrial Postgraduate Program with Nanyang Technological University (NTU). The calculations for this work were partially conducted using the resources of the National Supercomputing Center, Singapore and High-Performance Computing Center, NTU.

## Supporting Information

Additional figures including details such as the geometry, band structure, DOS, defect formation energy, spin density, potential, and charge density differences.

## References

1. Nakamura, S., Background Story of the Invention of Efficient Blue InGaN Light Emitting Diodes (Nobel Lecture). *Annalen der Physik*. **2015,** *527*, 335–349.
2. Huang, Y.; Duan, X.; Cui, Y.; Lieber, C. M., Gallium Nitride Nanowire Nanodevices. *Nano Letters*. **2002,** *2*, 101–104.
3. Palacios, T.; Chakraborty, A.; Heikman, S.; Keller, S.; DenBaars, S. P.; Mishra, U. K., AlGaN/GaN High Electron Mobility Transistors with InGaN Back-Barriers. *IEEE Electron Device Letters*. **2005,** *27*, 13–15.
4. Zhu, J.; Zhou, X.; Jing, L.; Hua, Q.; Hu, W.; Wang, Z. L., Piezotronic Effect Modulated Flexible AlGaN/GaN High-Electron-Mobility Transistors. *ACS Nano*. **2019,** *13*, 13161–13168.
5. Glavin, N. R.; Chabak, K. D.; Heller, E. R.; Moore, E. A.; Prusnick, T. A.; Maruyama, B.; Walker, D. E., Jr.; Dorsey, D. L.; Paduano, Q.; Snure, M., Flexible Gallium Nitride for High-Performance, Strainable Radio-Frequency Devices. *Advanced Materials*. **2017,** *29*, 1701838.
6. Dahal, R.; Li, J.; Aryal, K.; Lin, J. Y.; Jiang, H. X., InGaN/GaN Multiple Quantum Well Concentrator Solar Cells. *Applied Physics Letters*. **2010,** *97*, 073115.
7. Jiang, C.; Jing, L.; Huang, X.; Liu, M.; Du, C.; Liu, T.; Pu, X.; Hu, W.; Wang, Z. L., Enhanced Solar Cell Conversion Efficiency of InGaN/GaN Multiple Quantum Wells by Piezo-Phototronic Effect. *ACS Nano*. **2017,** *11*, 9405–9412.
8. Bae, H.; Rho, H.; Min, J.-W.; Lee, Y.-T.; Lee, S. H.; Fujii, K.; Lee, H.-J.; Ha, J.-S., Improvement of Efficiency in Graphene/Gallium Nitride Nanowire on Silicon Photoelectrode for Overall Water Splitting. *Applied Surface Science*. **2017,** *422*, 354–358.

9. Kibria, M. G.; Nguyen, H. P. T.; Cui, K.; Zhao, S.; Liu, D.; Guo, H.; Trudeau, M. L.; Paradis, S.; Hakima, A.-R.; Mi, Z., One-Step Overall Water Splitting under Visible Light Using Multiband InGaN/GaN Nanowire Heterostructures. *ACS Nano*. **2013,** *7*, 7886–7893.
10. Chen, Y.; Liu, K.; Liu, J.; Lv, T.; Wei, B.; Zhang, T.; Zeng, M.; Wang, Z.; Fu, L., Growth of 2D GaN Single Crystals on Liquid Metals. *Journal of the American Chemical Society*. **2018,** *140*, 16392–16395.
11. Qin, G.; Qin, Z.; Wang, H.; Hu, M., Anomalously Temperature-Dependent Thermal Conductivity of Monolayer GaN with Large Deviations from the Traditional 1/T Law. *Physical Review B*. **2017,** *95*, 195416.
12. Al Balushi, Z. Y.; Wang, K.; Ghosh, R. K.; Vila, R. A.; Eichfeld, S. M.; Caldwell, J. D.; Qin, X.; Lin, Y. C.; DeSario, P. A.; Stone, G.; Subramanian, S.; Paul, D. F.; Wallace, R. M.; Datta, S.; Redwing, J. M.; Robinson, J. A., Two-Dimensional Gallium Nitride Realized Via Graphene Encapsulation. *Nature Materials*. **2016,** *15*, 1166–1171.
13. Zhou, M.; Lu, Y. H.; Cai, Y. Q.; Zhang, C.; Feng, Y. P., Adsorption of Gas Molecules on Transition Metal Embedded Graphene: A Search for High-Performance Graphene-Based Catalysts and Gas Sensors. *Nanotechnology*. **2011,** *22*, 385502.
14. Zhang, Y. H.; Chen, Y. B.; Zhou, K. G.; Liu, C. H.; Zeng, J.; Zhang, H. L.; Peng, Y., Improving Gas Sensing Properties of Graphene by Introducing Dopants and Defects: A First-Principles Study. *Nanotechnology*. **2009,** *20*, 185504.
15. Hu, W.; Xia, N.; Wu, X.; Li, Z.; Yang, J., Silicene as a Highly Sensitive Molecule Sensor for $NH_3$, NO and $NO_2$. *Physical Chemistry Chemical Physics*. **2014,** *16*, 6957–6962.
16. Xia, W.; Hu, W.; Li, Z.; Yang, J., A First-Principles Study of Gas Adsorption on Germanene. *Physical Chemistry Chemical Physics*. **2014,** *16*, 22495–22498.
17. Cai, Y.; Zhang, G.; Zhang, Y.-W., Charge Transfer and Functionalization of Monolayer InSe by Physisorption of Small Molecules for Gas Sensing. *The Journal of Physical Chemistry C*. **2017,** *121*, 10182–10193.
18. Cui, Z.; Wang, X.; Ding, Y.; Li, E.; Bai, K.; Zheng, J.; Liu, T., Adsorption of CO, $NH_3$, NO, and $NO_2$ on Pristine and Defective G-GaN: Improved Gas Sensing and Functionalization. *Applied Surface Science*. **2020,** *530*, 147275.
19. Wu, P.; Li, Y.; Yang, A.; Tan, X.; Chu, J.; Zhang, Y.; Yan, Y.; Tang, J.; Yuan, H.; Zhang, X.; Xiao, S., Advances in 2D Materials Based Gas Sensors for Industrial Machine Olfactory Applications. *ACS Sensors*. **2024,** *9*, 2728–2776.
20. Hussain, T.; Kaewmaraya, T.; Chakraborty, S.; Ahuja, R., Defect and Substitution-Induced Silicene Sensor to Probe Toxic Gases. *The Journal of Physical Chemistry C*. **2016,** *120*, 25256–25262.
21. Cai, Y.; Ke, Q.; Zhang, G.; Zhang, Y.-W., Energetics, Charge Transfer, and Magnetism of Small Molecules Physisorbed on Phosphorene. *The Journal of Physical Chemistry C*. **2015,** *119*, 3102–3110.
22. Dingle, R.; Sell, D. D.; Stokowski, S. E.; Ilegems, M., Absorption, Reflectance, and Luminescence of GaN Epitaxial Layers. *Physical Review B*. **1971,** *4*, 1211–1218.
23. Qin, Z.; Qin, G.; Zuo, X.; Xiong, Z.; Hu, M., Orbitally Driven Low Thermal Conductivity of Monolayer Gallium Nitride (GaN) with Planar Honeycomb Structure: A Comparative Study. *Nanoscale*. **2017,** *9*, 4295–4309.
24. Şahin, H.; Cahangirov, S.; Topsakal, M.; Bekaroglu, E.; Akturk, E.; Senger, R. T.; Ciraci, S., Monolayer Honeycomb Structures of Group-IV Elements and III-V Binary Compounds: First-Principles Calculations. *Physical Review B*. **2009,** *80*, 155453.

25. Sun, C.; Yang, M.; Wang, T.; Shao, Y.; Wu, Y.; Hao, X., Graphene-Oxide-Assisted Synthesis of GaN Nanosheets as a New Anode Material for Lithium-Ion Battery. *ACS Applied Materials & Interfaces*. **2017,** *9*, 26631–26636.
26. Sreedhara, M. B.; Vasu, K.; Rao, C. N. R., Synthesis and Characterization of Few-Layer Nanosheets of GaN and Other Metal Nitrides. *Zeitschrift für anorganische und allgemeine Chemie*. **2014,** *640*, 2737–2741.
27. Rong, B.; Salemink, H. W.; Roeling, E. M.; van der Heijden, R.; Karouta, F.; van der Drift, E., Fabrication of Two Dimensional GaN Nanophotonic Crystals (31). *Journal of Vacuum Science & Technology B: Microelectronics and Nanometer Structures Processing, Measurement, and Phenomena*. **2007,** *25*, 2632–2636.
28. Sahu, T. K.; Sahu, S. P.; Hembram, K. P. S. S.; Lee, J.-K.; Biju, V.; Kumar, P., Free-Standing 2D Gallium Nitride for Electronic, Excitonic, Spintronic, Piezoelectric, Thermoplastic, and 6G Wireless Communication Applications. *NPG Asia Materials*. **2023,** *15*, 49.
29. Yazyev, O. V., Emergence of Magnetism in Graphene Materials and Nanostructures. *Reports on Progress in Physics*. **2010,** *73*, 056501.
30. Frey, N. C.; Bandyopadhyay, A.; Kumar, H.; Anasori, B.; Gogotsi, Y.; Shenoy, V. B., Surface-Engineered Mxenes: Electric Field Control of Magnetism and Enhanced Magnetic Anisotropy. *ACS Nano*. **2019,** *13*, 2831–2839.
31. Wang, Y.; Yi, J., Ferromagnetism in Two-Dimensional Materials Via Doping and Defect Engineering. In *Spintronic 2D Materials*, Elsevier: 2020; pp 95–124.
32. Lei, Z.; Sathish, C. I.; Geng, X.; Guan, X.; Liu, Y.; Wang, L.; Qiao, L.; Vinu, A.; Yi, J., Manipulation of Ferromagnetism in Intrinsic Two-Dimensional Magnetic and Nonmagnetic Materials. *Matter*. **2022,** *5*, 4212–4273.
33. Papavasileiou, A. V.; Menelaou, M.; Sarkar, K. J.; Sofer, Z.; Polavarapu, L.; Mourdikoudis, S., Ferromagnetic Elements in Two-Dimensional Materials: 2D Magnets and Beyond. *Advanced Functional Materials*. **2023,** *34*, 2309046.
34. Liang, Q.; Zhang, Q.; Zhao, X.; Liu, M.; Wee, A. T. S., Defect Engineering of Two-Dimensional Transition-Metal Dichalcogenides: Applications, Challenges, and Opportunities. *ACS Nano*. **2021,** *15*, 2165–2181.
35. Ying, Y.; Fan, K.; Lin, Z.; Huang, H., Facing the "Cutting Edge:" Edge Site Engineering on 2D Materials for Electrocatalysis and Photocatalysis. *Advanced Materials*. **2025,** *37*, e2418757.
36. Joshi, M.; Ren, X.; Lin, T.; Joshi, R., Mechanistic Insights into Gas Adsorption on 2D Materials. *Small*. **2025,** *21*, e2406706.
37. Iloanya, A. C.; Kastuar, S. M.; Jana, G.; Ekuma, C. E., Atomic-Scale Intercalation and Defect Engineering for Enhanced Magnetism and Optoelectronic Properties in Atomically Thin GeS. *Scientific Reports*. **2025,** *15*, 4546.
38. Zhao, Q.; Xiong, Z.; Qin, Z.; Chen, L.; Wu, N.; Li, X., Tuning Magnetism of Monolayer GaN by Vacancy and Nonmagnetic Chemical Doping. *Journal of Physics and Chemistry of Solids*. **2016,** *91*, 1–6.
39. Gao, H.; Ye, H.; Yu, Z.; Zhang, Y.; Liu, Y.; Li, Y., Point Defects and Composition in Hexagonal Group-III Nitride Monolayers: A First-Principles Calculation. *Superlattices and Microstructures*. **2017,** *112*, 136–142.
40. González, R.; López-Pérez, W.; González-García, Á.; Moreno-Armenta, M. G.; González-Hernández, R., Vacancy Charged Defects in Two-Dimensional GaN. *Applied Surface Science*. **2018,** *433*, 1049–1055.

41. Cui, Z.; Wang, X.; Li, E.; Ding, Y.; Sun, C.; Sun, M., Alkali-Metal-Adsorbed G-GaN Monolayer: Ultralow Work Functions and Optical Properties. *Nanoscale Research Letters*. **2018,** *13*, 207.
42. Singh, A. K.; Hennig, R. G., Computational Synthesis of Single-Layer GaN on Refractory Materials. *Applied Physics Letters*. **2014,** *105*, 051604.
43. Jia, Y.; Shi, Z.; Hou, W.; Zang, H.; Jiang, K.; Chen, Y.; Zhang, S.; Qi, Z.; Wu, T.; Sun, X.; Li, D., Elimination of the Internal Electrostatic Field in Two-Dimensional GaN-Based Semiconductors. *npj 2D Materials and Applications*. **2020,** *4*, 1–7.
44. Lu, H.; Guo, Y.; Robertson, J., Chemical Trends of Schottky Barrier Behavior on Monolayer Hexagonal B, Al, and Ga Nitrides. *Journal of Applied Physics*. **2016,** *120*, 065302.
45. Mu, Y., Chemical Functionalization of Gan Monolayer by Adatom Adsorption. *The Journal of Physical Chemistry C*. **2015,** *119*, 20911–20916.
46. Onen, A.; Kecik, D.; Durgun, E.; Ciraci, S., Gan: From Three- to Two-Dimensional Single-Layer Crystal and Its Multilayer Van Der Waals Solids. *Physical Review B*. **2016,** *93*, 085431.
47. Bikerouin, M.; Balli, M., Electric Field and Strain Induced Gap Modifications in Multilayered GaN. *Applied Surface Science*. **2022,** *578*, 151970.
48. Yao, W.; Wu, B.; Liu, Y., Growth and Grain Boundaries in 2D Materials. *ACS Nano*. **2020,** *14*, 9320–9346.
49. Zou, X.; Yakobson, B. I., An Open Canvas—2D Materials with Defects, Disorder, and Functionality. *Accounts of chemical research*. **2015,** *48*, 73–80.
50. Zheng, Y. J.; Chen, Y.; Huang, Y. L.; Gogoi, P. K.; Li, M.-Y.; Li, L.-J.; Trevisanutto, P. E.; Wang, Q.; Pennycook, S. J.; Wee, A. T. S.; Quek, S. Y., Point Defects and Localized Excitons in 2D $WSe_2$. *ACS Nano*. **2019,** *13*, 6050–6059.
51. Liu, Y.; Xiao, C.; Li, Z.; Xie, Y., Vacancy Engineering for Tuning Electron and Phonon Structures of Two-Dimensional Materials. *Advanced Energy Materials*. **2016,** *6*, 1600436.
52. Zhou, W.; Zou, X.; Najmaei, S.; Liu, Z.; Shi, Y.; Kong, J.; Lou, J.; Ajayan, P. M.; Yakobson, B. I.; Idrobo, J. C., Intrinsic Structural Defects in Monolayer Molybdenum Disulfide. *Nano Letters*. **2013,** *13*, 2615–2622.
53. Ghorbani-Asl, M.; Kretschmer, S.; Krasheninnikov, A. V., Two-Dimensional Materials under Ion Irradiation: From Defect Production to Structure and Property Engineering. In *Defects in Two-Dimensional Materials*, 2022; pp 259–301.
54. Jiang, J.; Xu, T.; Lu, J.; Sun, L.; Ni, Z., Defect Engineering in 2D Materials: Precise Manipulation and Improved Functionalities. *Research*. **2019,** *2019*, 4641739.
55. He, T.; Wang, Z.; Zhong, F.; Fang, H.; Wang, P.; Hu, W., Etching Techniques in 2D Materials. *Advanced Materials Technologies*. **2019,** *4*, 1900064
56. Rasool, H. I.; Ophus, C.; Zettl, A., Atomic Defects in Two Dimensional Materials. *Advanced Materials*. **2015,** *27*, 5771–5777.
57. Look, D. C.; Reynolds, D. C.; Hemsky, J. W.; Sizelove, J. R.; Jones, R. L.; Molnar, R. J., Defect Donor and Acceptor in GaN. *Physical Review Letters*. **1997,** *79*, 2273.
58. Nykänen, H.; Suihkonen, S.; Kilanski, L.; Sopanen, M.; Tuomisto, F., Low Energy Electron Beam Induced Vacancy Activation in GaN. *Applied Physics Letters*. **2012,** *100*, 122105.
59. Zhao, Y.; Tripathi, M.; Cernevics, K.; Avsar, A.; Ji, H. G.; Gonzalez Marin, J. F.; Cheon, C. Y.; Wang, Z.; Yazyev, O. V.; Kis, A., Electrical Spectroscopy of Defect States and Their Hybridization in Monolayer $MoS_2$. *Nature Communications*. **2023,** *14*, 44.
60. Wan, Y.; Li, E.; Yu, Z.; Huang, J. K.; Li, M. Y.; Chou, A. S.; Lee, Y. T.; Lee, C. J.; Hsu, H. C.; Zhan, Q.; Aljarb, A.; Fu, J. H.; Chiu, S. P.; Wang, X.; Lin, J. J.; Chiu, Y. P.; Chang,

W. H.; Wang, H.; Shi, Y.; Lin, N., *et al.*, Low-Defect-Density $WS_2$ by Hydroxide Vapor Phase Deposition. *Nature Communications*. **2022,** *13*, 4149.
61. Feng, S.; Tan, J.; Zhao, S.; Zhang, S.; Khan, U.; Tang, L.; Zou, X.; Lin, J.; Cheng, H. M.; Liu, B., Synthesis of Ultrahigh-Quality Monolayer Molybdenum Disulfide through in Situ Defect Healing with Thiol Molecules. *Small*. **2020,** *16*, e2003357.
62. Uedono, A.; Tanaka, R.; Takashima, S.; Ueno, K.; Edo, M.; Shima, K.; Chichibu, S. F.; Uzuhashi, J.; Ohkubo, T.; Ishibashi, S.; Sierakowski, K.; Bockowski, M., Vacancy-Type Defects and Their Trapping/Detrapping of Charge Carriers in Ion-Implanted GaN Studied by Positron Annihilation. *physica status solidi (b)*. **2024,** *261*, 2400060.
63. Ma, Y.; Dai, Y.; Guo, M.; Niu, C.; Yu, L.; Huang, B., Magnetic Properties of the Semifluorinated and Semihydrogenated 2D Sheets of Group-IV and III-V Binary Compounds. *Applied Surface Science*. **2011,** *257*, 7845–7850.
64. Diallo, I. C.; Demchenko, D. O., Native Point Defects in GaN: A Hybrid-Functional Study. *Physical Review Applied*. **2016,** *6*, 064002.
65. Jia, W.; Niu, Y.; Zhou, M.; Liu, R.; Zhang, L.; Wang, X.; Ji, W., Effects of Vacancy Defects on the Electronic Structure and Optical Properties of GaN:Fe. *Superlattices and Microstructures*. **2019,** *133*, 106152.
66. Arrhenius, S., Über Die Dissociationswärme Und Den Einfluss Der Temperatur Auf Den Dissociationsgrad Der Elektrolyte. *Zeitschrift für physikalische Chemie*. **1889,** *4*, 96–116.
67. Peng, S.; Cho, K.; Qi, P.; Dai, H., Ab Initio Study of CNT $NO_2$ Gas Sensor. *Chemical Physics Letters*. **2004,** *387*, 271–276.
68. Nath, U.; Sarma, M., Realization of Efficient and Selective NO and $NO_2$ Detection Via Surface Functionalized H-$B_2S_2$ Monolayer. *Physical Chemistry Chemical Physics*. **2024,** *26*, 12386–12396.
69. Cui, H.; Zhang, X.; Zhang, G.; Tang, J., Pd-Doped $MoS_2$ Monolayer: A Promising Candidate for DGA in Transformer Oil Based on DFT Method. *Applied Surface Science*. **2019,** *470*, 1035–1042.
70. Hajati, Y.; Blom, T.; Jafri, S. H.; Haldar, S.; Bhandary, S.; Shoushtari, M. Z.; Eriksson, O.; Sanyal, B.; Leifer, K., Improved Gas Sensing Activity in Structurally Defected Bilayer Graphene. *Nanotechnology*. **2012,** *23*, 505501.
71. Mehdi Aghaei, S.; Monshi, M. M.; Torres, I.; Zeidi, S. M. J.; Calizo, I., DFT Study of Adsorption Behavior of NO, CO, $NO_2$, and $NH_3$ Molecules on Graphene-Like $BC_3$: A Search for Highly Sensitive Molecular Sensor. *Applied Surface Science*. **2018,** *427*, 326–333.
72. Marjani, A.; Ghashghaee, M.; Ghambarian, M.; Ghadiri, M., Scandium Doping of Black Phosphorene for Enhanced Sensitivity to Hydrogen Sulfide: Periodic DFT Calculations. *Journal of Physics and Chemistry of Solids*. **2021,** *148*, 109765.
73. Kittel, C., *Introduction to Solid State Physics*. John Wiley & Sons, Inc.: New York, 1953.
74. Li, F.; Shi, C., No-Sensing Performance of Vacancy Defective Monolayer $MoS_2$ Predicted by Density Function Theory. *Applied Surface Science*. **2018,** *434*, 294–306.
75. Liu, F.; Zhou, J.; Zhu, C.; Liu, Z., Electric Field Effect in Two-Dimensional Transition Metal Dichalcogenides. *Advanced Functional Materials*. **2016,** *27*, 1602404.
76. Ramasubramaniam, A.; Naveh, D.; Towe, E., Tunable Band Gaps in Bilayer Transition-Metal Dichalcogenides. *Physical Review B*. **2011,** *84*, 205325.
77. Yue, Q.; Shao, Z.; Chang, S.; Li, J., Adsorption of Gas Molecules on Monolayer $MoS_2$ and Effect of Applied Electric Field. *Nanoscale Research Letters*. **2013,** *8*, 1–7.

78. Dobrinsky, A.; Simin, G.; Gaska, R.; Shur, M., III-Nitride Materials and Devices for Power Electronics. *ECS Transactions*. **2013,** *58*, 129–143.
79. Chow, T.; Ghezzo, M., SiC Power Devices. *MRS Online Proceedings Library*. **1996,** *423*, 9–21.
80. Li, Y.; Xia, L.; Li, N.; Tang, S.; Ge, Y.; Wang, J.; Xiao, B.; Cheng, Y.; Ang, L. K.; Meng, G., Uncovering a Widely Applicable Empirical Formula for Field Emission Characteristics of Metallic Nanotips in Nanogaps. *Nature Communications*. **2025,** *16*, 5583.
81. Weintrub, B. I.; Hsieh, Y. L.; Kovalchuk, S.; Kirchhof, J. N.; Greben, K.; Bolotin, K. I., Generating Intense Electric Fields in 2D Materials by Dual Ionic Gating. *Nature Communications*. **2022,** *13*, 6601.
82. Sanders, N.; Bayerl, D.; Shi, G.; Mengle, K. A.; Kioupakis, E., Electronic and Optical Properties of Two-Dimensional GaN from First-Principles. *Nano Letters*. **2017,** *17*, 7345–7349.
83. Bekaroglu, E.; Topsakal, M.; Cahangirov, S.; Ciraci, S., First-Principles Study of Defects and Adatoms in Silicon Carbide Honeycomb Structures. *Physical Review B*. **2010,** *81*, 075433.
84. Huang, L.; Liu, H.; Deng, X.; Cui, W., The Structural, Mechanical and Electrical Properties of 2D SiC with C-Related Point Defects and Substitution of C by Foreign Atoms. *Vacuum*. **2023,** *208*, 111700.
85. Xu, Z.; Li, Y.; Liu, Z.; Li, C., Dependence of Electronic and Optical Properties of Multilayer SiC and GeC on Stacking Sequence and External Electric Field. *Physica E: Low-dimensional Systems and Nanostructures*. **2016,** *79*, 198–205.
86. Mohseni, M.; Sarsari, I. A.; Karbasizadeh, S.; Hassanzada, Q.; Ala-Nissila, T.; Gali, A., Vacancy-Related Color Centers in Two-Dimensional Silicon Carbide Monolayers. *Physical Review Materials*. **2024,** *8*, 056201.
87. Polo, V.; Kraka, E.; Cremer, D., Electron Correlation and the Self-Interaction Error of Density Functional Theory. *Molecular Physics*. **2002,** *100*, 1771–1790.
88. Lyons, J. L.; Janotti, A.; Van de Walle, C. G., Effects of Hole Localization on Limiting P-Type Conductivity in Oxide and Nitride Semiconductors. *Journal of Applied Physics*. **2014,** *115*, 012014.
89. Wellendorff, J.; Lundgaard, K. T.; Møgelhøj, A.; Petzold, V.; Landis, D. D.; Nørskov, J. K.; Bligaard, T.; Jacobsen, K. W., Density Functionals for Surface Science: Exchange-Correlation Model Development with Bayesian Error Estimation. *Physical Review B*. **2012,** *85*, 235149.
90. Krukau, A. V.; Vydrov, O. A.; Izmaylov, A. F.; Scuseria, G. E., Influence of the Exchange Screening Parameter on the Performance of Screened Hybrid Functionals. *The Journal of Chemical Physics*. **2006,** *125*, 224106.
91. Flores, M. A.; Orellana, W.; Menéndez-Proupin, E., Accuracy of the Heyd-Scuseria-Ernzerhof Hybrid Functional to Describe Many-Electron Interactions and Charge Localization in Semiconductors. *Physical Review B*. **2018,** *98*, 155131.
92. Schimka, L.; Harl, J.; Kresse, G., Improved Hybrid Functional for Solids: The Hsesol Functional. *The Journal of Chemical Physics*. **2011,** *134*, 024116.
93. Perdew, J. P.; Burke, K.; Ernzerhof, M., Generalized Gradient Approximation Made Simple. *Physical Review Letters*. **1996,** *77*, 3865.
94. Grimme, S., Semiempirical Gga-Type Density Functional Constructed with a Long-Range Dispersion Correction. *Journal of Computational Chemistry*. **2006,** *27*, 1787–1799.
95. Kanwal, A.; Jalil, A.; Ilyas, S. Z.; Ahmed, S.; Agathopoulos, S.; Znaidia, S., Effect of Electric Field on Two-Dimensional Honeycomb Structures from Group (III–V). *Journal of Physics and Chemistry of Solids*. **2022,** *162*, 110507.

96. Wang, V.; Xu, N.; Liu, J.-C.; Tang, G.; Geng, W.-T., Vaspkit: A User-Friendly Interface Facilitating High-Throughput Computing and Analysis Using VASP Code. *Computer Physics Communications*. **2021,** *267*, 108033.
97. Henkelman, G.; Arnaldsson, A.; Jónsson, H., A Fast and Robust Algorithm for Bader Decomposition of Charge Density. *Computational Materials Science*. **2006,** *36*, 354–360.
98. Sanville, E.; Kenny, S. D.; Smith, R.; Henkelman, G., Improved Grid-Based Algorithm for Bader Charge Allocation. *Journal of Computational Chemistry*. **2007,** *28*, 899–908.
99. Tang, W.; Sanville, E.; Henkelman, G., A Grid-Based Bader Analysis Algorithm without Lattice Bias. *Journal of Physics: Condensed Matter*. **2009,** *21*, 084204.
100. Radin, M. *VASPLAB*, 1.1.1.0; 2024.
101. Momma, K.; Izumi, F., VESTA 3 for Three-Dimensional Visualization of Crystal, Volumetric and Morphology Data. *Journal of Applied Crystallography*. **2011,** *44*, 1272–1276.
102. Togo, A.; Chaput, L.; Tadano, T.; Tanaka, I., Implementation Strategies in Phonopy and Phono3py. *Journal of Physics: Condensed Matter*. **2023,** *35*, 353001.
103. Togo, A., First-Principles Phonon Calculations with Phonopy and Phono3py. *Journal of the Physical Society of Japan*. **2022,** *92*, 012001.

**TOC**

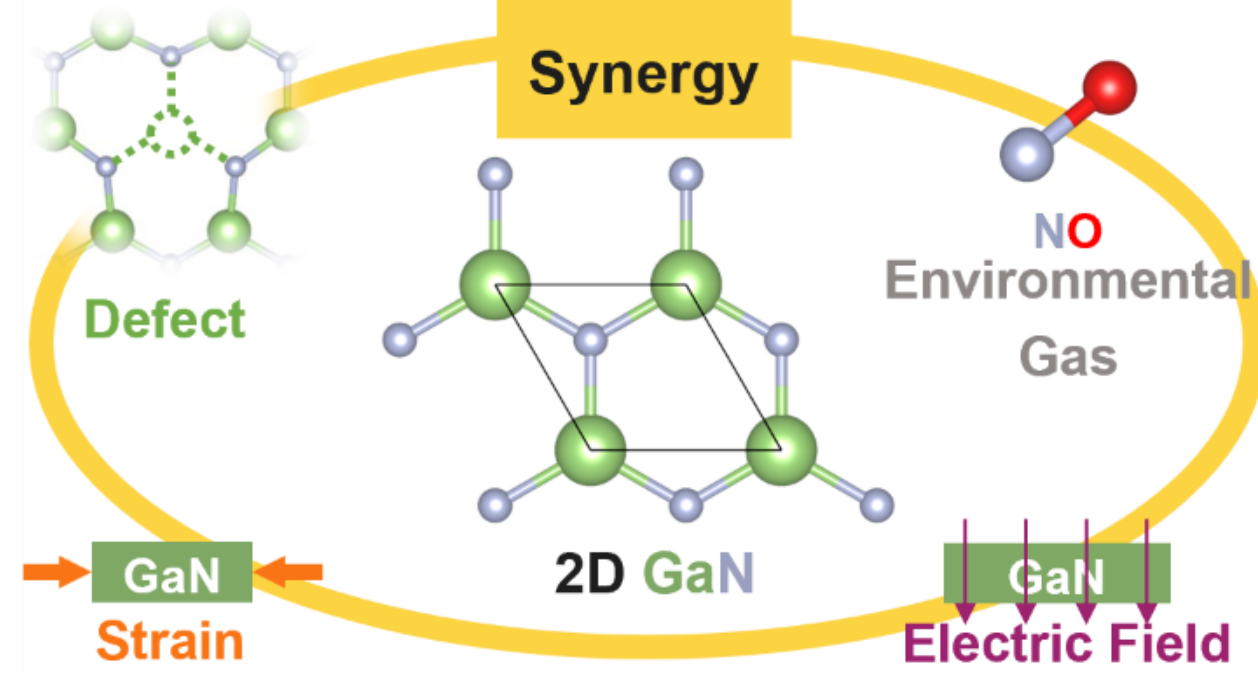